\documentclass[
 reprint,
 superscriptaddress,
 preprintnumbers,
 amsmath
 amssymb,
 prx,
 floatfix,
]{revtex4-2}

\DeclareMathAlphabet{\mathsfit}{T1}{\sfdefault}{\mddefault}{\sldefault}
\SetMathAlphabet{\mathsfit}{bold}{T1}{\sfdefault}{\bfdefault}{\sldefault}

\usepackage{amsmath}
\usepackage{graphicx}% Include figure files
\graphicspath{{fig/}}   %add figure path
\usepackage{dcolumn}% Align table columns on decimal point
\usepackage{bm}% bold math
\usepackage{makecell}
\usepackage{braket} %,mleftright}
\usepackage{siunitx}
\usepackage{color}
\usepackage{array}
\usepackage{xr}
\usepackage{upgreek}
\usepackage[colorlinks]{hyperref}
\usepackage[capitalize]{cleveref}
\usepackage[caption=false]{subfig}
\usepackage{multirow} % for tables

\newcommand{\supp}{Supplementary Information}

\begin{document} 

\title{Implementing a Synthetic Magnetic Vector Potential\\in a 2D Superconducting Qubit Array}

\def\RLEaffil{Research Laboratory of Electronics, Massachusetts Institute of Technology, Cambridge, MA 02139, USA}
\def\LLaffil{MIT Lincoln Laboratory, Lexington, MA 02421, USA}
\def\Physaffil{Department of Physics, Massachusetts Institute of Technology, Cambridge, MA 02139, USA}
\def\EECSaffil{Department of Electrical Engineering and Computer Science, Massachusetts Institute of Technology, Cambridge, MA 02139, USA}

\author{Ilan~T.~Rosen}
\email{itrosen@mit.edu}
\affiliation{\RLEaffil}

\author{Sarah~Muschinske}
\affiliation{\RLEaffil}
\affiliation{\EECSaffil}

\author{Cora~N.~Barrett}
\affiliation{\RLEaffil} 
\affiliation{\Physaffil}

\author{Arkya~Chatterjee}
\affiliation{\Physaffil}

\author{Max~Hays}
\affiliation{\RLEaffil}

\author{Michael~A.~DeMarco}
\affiliation{\Physaffil}

\author{Amir~H.~Karamlou}
\affiliation{\RLEaffil}
\affiliation{\EECSaffil}

\author{David~A.~Rower}
\affiliation{\RLEaffil}
\affiliation{\Physaffil}

\author{Rabindra~Das}
\affiliation{\LLaffil}

\author{David~K.~Kim}
\affiliation{\LLaffil}

\author{Bethany~M.~Niedzielski}
\affiliation{\LLaffil}

\author{Meghan~Schuldt}
\affiliation{\LLaffil}

\author{Kyle~Serniak}
\affiliation{\RLEaffil}
\affiliation{\LLaffil}

\author{Mollie~E.~Schwartz}
\affiliation{\LLaffil}

\author{Jonilyn~L.~Yoder}
\affiliation{\LLaffil}

\author{Jeffrey~A.~Grover}
\affiliation{\RLEaffil}

\author{William~D.~Oliver}
\email{william.oliver@mit.edu}
\affiliation{\RLEaffil}
\affiliation{\EECSaffil}
\affiliation{\Physaffil}

\date{\today}

\begin{abstract}
Superconducting quantum processors are a compelling platform for analog quantum simulation due to the precision control, fast operation, and site-resolved readout inherent to the hardware.
Arrays of coupled superconducting qubits natively emulate the dynamics of interacting particles according to the Bose-Hubbard model.
However, many interesting condensed-matter phenomena emerge only in the presence of electromagnetic fields.
Here, we emulate the dynamics of charged particles in an electromagnetic field using a superconducting quantum simulator.
We realize a broadly adjustable synthetic magnetic vector potential by applying continuous modulation tones to all qubits.
We verify that the synthetic vector potential obeys requisite properties of electromagnetism: a spatially-varying vector potential breaks time-reversal symmetry and generates a gauge-invariant synthetic magnetic field, and a temporally-varying vector potential produces a synthetic electric field.
We demonstrate that the Hall effect---the transverse deflection of a charged particle propagating in an electromagnetic field---exists in the presence of the synthetic electromagnetic field. 
\end{abstract}

\maketitle

%+++++++++++++++++++++++++++++++++++++++++++++++++++++++++++++++++++++++++++++++++++++++++

\section{Introduction}
Analog quantum simulators emulate physical models of materials systems to elucidate their properties~\cite{greiner2002, houck2012, georgescu2014, scholl2021, altman2021, ebadi2021}.
At scale, these devices will operate in a regime that is intractable for classical computers, possibly providing scientific utility before quantum error correction enables general-purpose digital simulation~\cite{Preskill2018, kandala2019, daley2022, kim2023}.
Ideally, the hardware should be able to emulate models describing a variety of materials systems.
One important class of models describes electronic systems in materials where time-reversal symmetry is broken by a magnetic field. 
Magnetic fields are required to access many quantum phases of matter,
% particularly certain topological states including various quantum Hall states and Majorana systems~\cite{schnyder2008, qi2011, alicea2012, klitzing2020}, 
for example, certain quantum Hall states and systems supporting Majorana excitations~\cite{schnyder2008, qi2011, alicea2012, klitzing2020},
and are needed to observe several features of electrical transport such as the Aharonov-Bohm and Little-Parks effects~\cite{Aharonov1959, little1962}, quantum corrections to conductivity~\cite{hikami1980}, and aspects of electron optics~\cite{taychatanapat2013}.

Arrays of coupled superconducting qubits natively emulate tight-binding models.
The qubits represent the lattice sites, qubit excitations correspond to bosonic particles, and nearest-neighbor exchange interactions are realized by resonantly coupling adjacent qubits~\cite{ma2019, yan2019, yanay2020, saxberg2022, karamlou2022, karamlou2024}.
The ensuing dynamics respect time-reversal symmetry and therefore cannot emulate an electronic system in a magnetic field without adding a symmetry-breaking mechanism.
Although applying an external magnetic field generally breaks time-reversal symmetry for systems with charged particles, here such an emulator would retain time-reversal symmetry since the qubit excitations are chargeless bosons.

In the absence of a physical magnetic field, one can alternatively use the Harper-Hofstadter (HH) model,
\begin{equation}\label{eq:HH}
    \hat{H}_\mathrm{HH}/\hbar = \sum_{\langle i, j \rangle} J \left(e^{-i\phi_{ij}}\hat{a}_i^\dag \hat{a}_j + e^{i\phi_{ij}}\hat{a}_i \hat{a}_j^\dag \right),
\end{equation}
to emulate the dynamics of charged particles moving in a two-dimensional (2D) lattice in the presence of a perpendicular magnetic field~\cite{harper1955a, hofstadter1976}. 
Emulation of the HH model with uniform fields has been demonstrated with atomic simulators, where laser-assisted tunneling in a tilted optical lattice was used to set the phases and thereby break time-reversal symmetry~\cite{aidelsburger2013, miyake2013, leonard2023},
and in an array of microwave cavities, where ferrimagnetic inserts placed in certain cavities set the phases~\cite{owens2022}. 
Superconducting simulators comprising 1D qubit arrays have realized the Aubry-Andr\'e-Harper model, a related model that mimics the HH spectrum but not its dynamics as time-reversal symmetry is unbroken~\cite{aubry1980, roushan2017butterfly, li2023, xiang2023}.
Realizing the HH model with individually and dynamically adjustable phases $\phi_{ij}$ would enable a generalized emulator for particles in magnetic and electric fields via a (time-dependent) magnetic vector potential.

\begin{figure*}[ht!]
\subfloat{\label{fig:chip}}
\subfloat{\label{fig:flipchip}}
\subfloat{\label{fig:circuit}}
\subfloat{\label{fig:resonant}}
\subfloat{\label{fig:parametric}}
\subfloat{\label{fig:hopping_rate}}
\subfloat{\label{fig:hopping_amp}}
\includegraphics{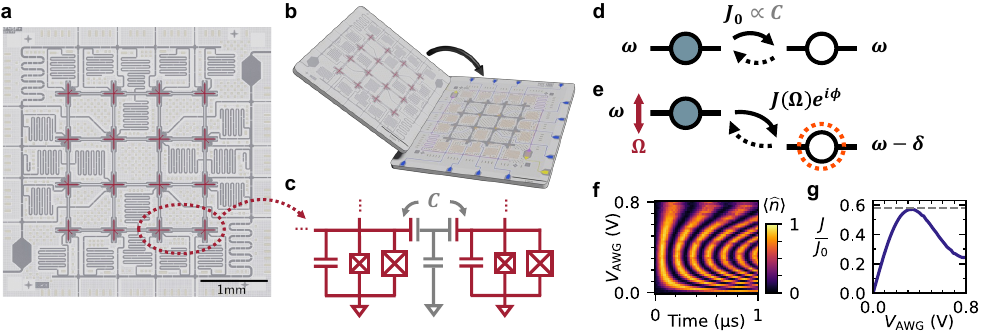}
\caption{\textbf{Generating Peierls phases using parametric coupling in a 16-qubit superconducting processor}. \textbf{(a)}~Optical micrograph of the superconducting quantum processor: a $4\times 4$ array of transmon qubits (red false color) with nearest-neighbor capacitive coupling. Adapted from Ref.~\cite{karamlou2024}. \textbf{(b)} Control and measurement lines are located on a separate chip situated in a flip-chip configuration. \textbf{(c)} Circuit diagram of two sites. Each site is a flux-tunable transmon qubit. Nearest-neighbor exchange coupling is mediated by a fixed mutual capacitance $C$. \textbf{(d)} Schematic of resonant coupling between two qubits, which is typically used in analog simulation experiments. Neighboring sites are brought to energetic resonance (vertical axis). A particle initialized on one site (teal filled circle at left) will hop to the other site (open circle at right) at the bare hopping rate $J_0$ determined by the mutual capacitance between the two sites. \textbf{(e)} Schematic of parametric coupling between two qubits, which is used in the present experiment. Neighboring sites are energetically detuned by $\hbar \delta$, and the energy of one site is modulated at frequency $\delta$ using the site's flux control line. The hopping rate depends on the amplitude $\Omega$ of the modulation. \textbf{(f)} Experimental demonstration of parametric hopping between two qubits. The population $\langle \hat{n} \rangle$ of the site initially unoccupied (orange circle in \textbf{(e)}) is shown versus time and the amplitude $V_\mathrm{AWG}$ of the signal applied to the flux line, which is proportional to $\Omega$. \textbf{(g)} The hopping rate normalized by the bare coupling strength $J_0$ as a function of $V_\mathrm{AWG}$, determined by fitting the data in \textbf{(f)}. The maximum value of the first Bessel function is indicated by the grey dashed line.} 
\label{fig:fig1}
\end{figure*}

In this work, we directly emulate the HH model using a 2D array of superconducting qubits by applying several control tones that together break time-reversal symmetry.
Our system comprises 16 transmon qubits~\cite{koch2007} with fixed capacitive coupling between adjacent qubits.
Instead of resonantly coupling the qubits, we detune adjacent qubits from one another and parametrically induce exchange interactions by modulating the qubits with control tones matching the detuning frequencies~\cite{alaeian2019, zhao2022}.
The phases of the modulation tones constitute a synthetic vector potential.
When the modulation tones have nonzero relative phases, time-reversal symmetry is broken and the processor adopts a synthetic perpendicular magnetic field.
When the phases vary in time, spatial-inversion symmetry is broken and the processor adopts a synthetic electric field according to Faraday's law of induction~\cite{lin2011}.
A broad range of electromagnetic field strengths and profiles, including spatially nonuniform and time-varying fields, can be emulated.

Using this emulator, we replicate several hallmark features of two-dimensional electronic systems in the presence of electromagnetic fields. 
We observe Aharonov-Bohm interference in rings of various lengths and verify the interference patterns are invariant to the choice of gauge.
We show that particles in a synthetic electric field exhibit Bloch oscillations rather than uniform motion.
Finally, upon simultaneous application of synthetic electric and magnetic fields, we observe the Hall effect: a propagating particle is deflected transversely to both fields.

\section{Methods}

\begin{figure*}[ht!]
\subfloat{\label{fig:2x2}}
\subfloat{\label{fig:3x3}}
\subfloat{\label{fig:4x4}}
\includegraphics{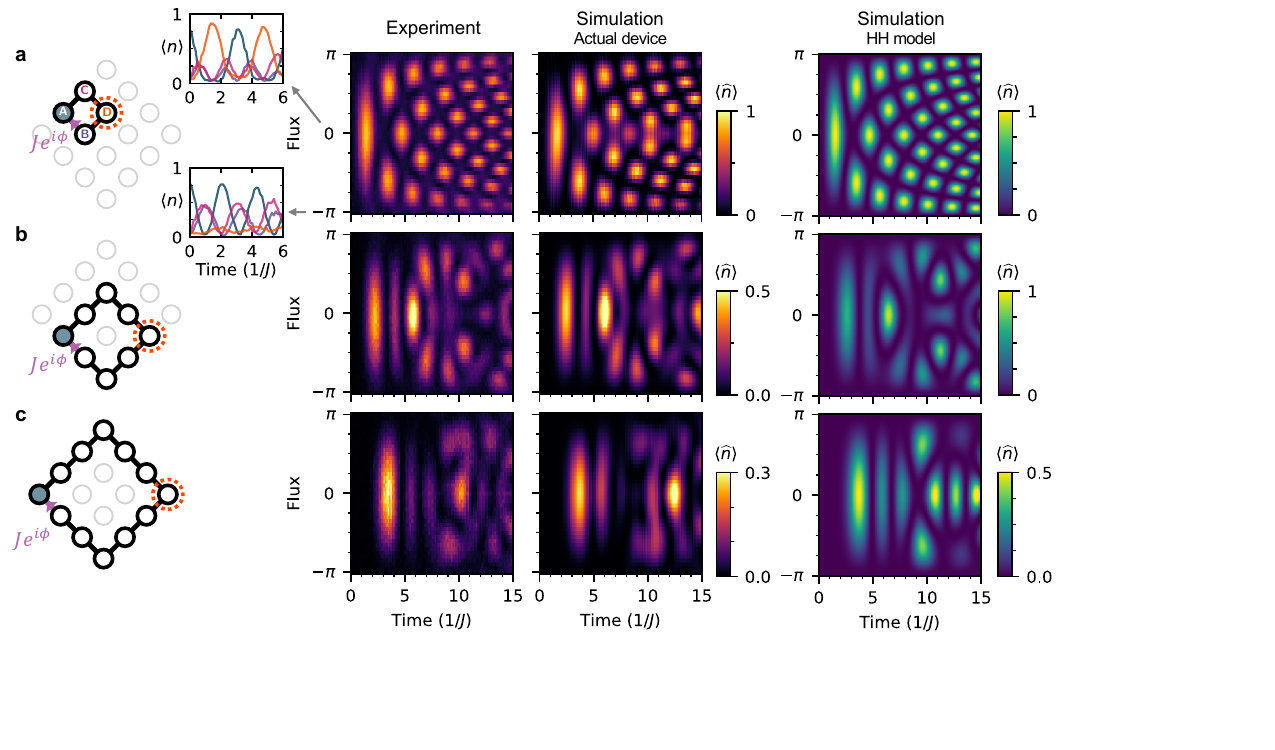}
\caption{\textbf{Aharonov-Bohm interference from a synthetic magnetic field}. A particle is initialized in the left corner of a ring (teal shaded site), and the population in the opposing corner (orange dashed circle) is presented as a function of time and the modulation phase of the purple highlighted exchange coupling. Interference patterns are shown in \textbf{(a)} a $2\times 2$ plaquette, \textbf{(b)} an $8$-site ring, and \textbf{(c)} a $12$-site ring. Experimental data are accompanied by the simulated operation of the superconducting processor (center column) along with simulations of the idealized HH model (rightmost column; note changing colorscale for visibility). Inset in \textbf{(a)}: time dynamics at (upper) $0$ and (lower) $-\pi$ phase. The populations of the four qubits in the plaquette are displayed in colors matching the site labels in the diagram at left.} 
\label{fig:fig2}
\end{figure*}

Our experiment is implemented on a superconducting quantum processor comprising 16 flux-tunable transmon qubits arranged in a $4\times 4$ grid (Fig.~\ref{fig:chip}).
Individual flux control lines and readout resonators are located on a separate chip arranged and brought in proximity to the qubits using a flip-chip configuration (Fig.~\ref{fig:flipchip})~\cite{rosenberg2017}.
The processor is discussed in more detail in Refs.~\cite{barrett2023, karamlou2024}. Nearest neighbors are transversely coupled through a fixed mutual capacitance (Fig.~\ref{fig:circuit}), realizing---within the rotating-wave approximation---the Bose-Hubbard model
\begin{align}\label{eq:BH}
    \hat{H}_\mathrm{BH}/\hbar &= \sum_i \left(\omega_i \hat{n}_i + \frac{U_i}{2} \hat{n}_i (\hat{n}_i-1) \right) \nonumber\\
    &+\sum_{\langle i, j \rangle} J_0^{ij} \left(\hat{a}_i^\dag \hat{a}_j+\hat{a}_i \hat{a}_j^\dag \right), %\frac{\hat{H}_J}{\hbar},
\end{align}
where $\hat{n}_i$ is the number operator corresponding to the bosonic annihilation operator $\hat{a}_i$ for a particle on site $i$, and the second summation extends over the $24$~nearest-neighbor pairs in the lattice. 
% Adjacent sites are coupled by the particle exchange interactions 
% \begin{equation}\label{eq:coupling}
%     \hat{H}_J/\hbar=\sum_{\langle i, j \rangle} J_0^{ij} \left(\hat{a}_i^\dag \hat{a}_j+\hat{a}_j^\dag \hat{a}_i \right),
% \end{equation}
% where the summation extends over the $24$~nearest-neighbor pairs in the lattice.
Here, a particle on site $i$ corresponds to an excitation of qubit $i$, and we refer to $\langle \hat{n}_i \rangle$ as its population.
The on-site energies $\omega_i$ are individually adjustable through the flux control lines.
The on-site interactions $U_i$ arise from the qubit anharmonicities and have an average strength $U/2\pi=\SI{-218(6)}{MHz}$, and the bare particle exchange interactions have real coefficients $J_0^{ij}$ with an average strength $J_0/2\pi=\SI{5.9(4)}{MHz}$.
For the remainder of this work, we consider only the single-particle manifold of Eq.~(\ref{eq:BH}) and therefore omit the on-site interaction term for brevity.

Typically, superconducting analog simulators operate with all sites resonant (uniform $\omega_i$) so that particles hop between sites at rate $J_0$ (Fig.~\ref{fig:resonant}). 
Such simulators respect time-reversal symmetry and emulate materials without a magnetic field. To emulate the dynamics of charged particles in an electromagnetic field, we wish to create a synthetic vector potential $\bm{A}$ that generates analogous dynamics for photons in the lattice. 
Through a Peierls substitution~\cite{peierls1933}, $\bm{A}$ maps to the Bose-Hubbard lattice as complex coefficients on the exchange terms. 
Known as Peierls phases, the complex phases are equivalent to $\bm{A}$ written in dimensionless units (a derivation is provided in Section~S1 of the \supp{}). 
To realize nonzero Peierls phases on each nearest-neighbor interaction, we employ a parametric coupling scheme, shown schematically in Fig.~\ref{fig:parametric}. 
Consider two neighboring qubits $i$ and $j$.
We adjust their frequencies to establish a detuning $\delta_{i}=\omega_i-\omega_j$ and sinusoidally modulate the on-site energy of qubit $i$ with frequency $\delta_{i}$, realizing the lab-frame Hamiltonian
\begin{align}\label{eq:lab}
    \hat{H}_L^{ij}/\hbar&=\Big(\omega_i + \Omega_i\sin(\delta_{i} t+\phi_{ij})\Big)\hat{n}_i + (\omega_i-\delta_{i})\hat{n}_j \nonumber\\
    &+J_0^{ij} \left(\hat{a}_i^\dag \hat{a}_j+\hat{a}_i \hat{a}_j^\dag \right),%\frac{H_J^{ij}}{\hbar},
\end{align}
where $\Omega_i$ is the modulation amplitude of qubit~$i$ and $\phi_{ij}$ is the Peierls phase.

Transforming Eq.~(\ref{eq:lab}) into the instantaneous rotating frame of both sites yields a Hamiltonian containing stationary terms
\begin{equation}\label{eq:rot}
    \hat{H}_R^{ij}/\hbar = J_0^{ij} \mathcal{J}_1\left(\frac{\Omega_i}{\delta_i}\right)\left(e^{-i\phi_{ij}}\hat{a}_i^\dag \hat{a}_j + e^{i\phi_{ij}}\hat{a}_i \hat{a}_j^\dag \right),
\end{equation}
where $\mathcal{J}_1(x)$ is the first-order Bessel function of the first kind, revealing that parametric modulations induce complex hopping.
The Peierls phase is the phase of the modulation $\phi_{ij}$.
Additional terms rotating at multiples of $\delta_i$ are discussed in the \supp{}.
We demonstrate experimentally that the hopping rate $J^{ij}=J_0^{ij} \mathcal{J}_1\left(\Omega_i/\delta_i\right)$ can be tuned by varying $\Omega_i$, reaching a maximum value $J^{ij}/J_0^{ij}\approx 0.58$ (Fig.~\ref{fig:hopping_rate},~\ref{fig:hopping_amp}), the maximum of $\mathcal{J}_1$.

Extending the parametric coupling throughout the qubit array emulates the HH model.
We note that, in our coupling scheme, the second qubit $j$ is also modulated at $\delta_j$ to induce parametric hopping to subsequent lattice sites, and so on. 
Choosing unequal values $\delta_{i}\neq \delta_{j}$ for all neighboring sites $i$,~$j$ enables full tunability of the synthetic vector potential; the layout of detunings and modulations is detailed in the~\supp{}. 
For the remainder of this work, the modulation of each qubit $i$ is chosen to provide a hopping rate $J^{ij}/2\pi=\SI{2.5}{MHz}$ to its neighbor $j$ when the latter is not modulated. 
The hopping rate $J^{ij}$ is reduced by a factor of approximately~$1.25$ when site $j$ is also modulated (see the \supp{}). 
The effective hopping rates are therefore approximately $J/2\pi\approx\SI{2.0}{MHz}$, such that $J/J_0\approx0.34$.

Secondly, we note that in Eq.~(\ref{eq:rot}), we present the parametric modulation as a sinusoid in the qubit frequency. 
In practice, we modulate the qubits sinusoidally in flux bias. 
Because the qubits are operated in the linear regime of their spectra, the two types of modulation are equivalent to first order and differences are insignificant at the utilized modulation amplitudes.
Thirdly, we note that $\delta_i$ may be positive or negative; both signs are used in the experiment. 
When $\delta_i<0$, the sign of the Peierls phase is opposite that of the modulation phase.
Lastly, the experimental data presented in this work are based on simultaneous single-shot population measurements of all active qubits.
Before determining the population on each site by averaging single-shot results, the single-shot data are post-selected on total population to partially mitigate relaxation and readout infidelity.

\section{Results and Discussion}

\begin{figure}[ht!]
\subfloat{\label{fig:phase7}}
\subfloat{\label{fig:phase3qb}}
\includegraphics{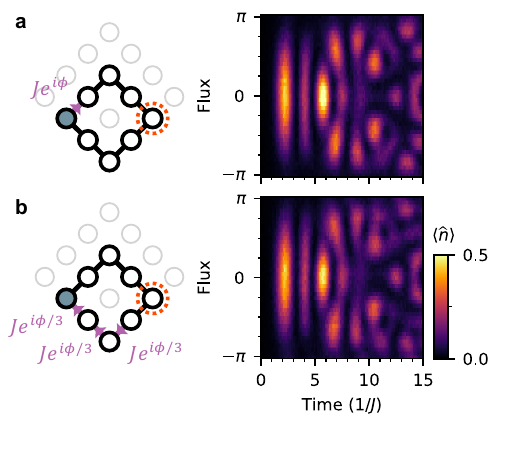}
\caption{\textbf{Gauge invariance.} Aharonov-Bohm interference in the same $8$-site ring as shown in Fig.~\ref{fig:3x3}, but \textbf{(a)} with the Peierls phase added to an exchange coupling on the upper, rather than lower, trajectory around the ring, and \textbf{(b)} with the phase $\phi$ accumulated through the lower trajectory distributed between Peierls phases $\phi/3$ added to three exchange couplings. Approximately matching interference patterns are observed in all cases, demonstrating that the dynamics only depends on the net flux through the ring and reflecting the gauge invariance of the synthetic magnetic field. Teal shading indicates the initial position of the particle, dashed orange circles indicate the site whose population is shown.} 
\label{fig:fig3}
\end{figure}

%+++++++++++++++++++ 2x2, 0 vs pi flux ++++++++++++++++++

The dynamics of a particle hopping between two sites alone does not depend on the Peierls phase of the intervening exchange coupling.
To observe Peierls phase-dependent dynamics, we need to consider a set of sites forming a closed path, in which case the dynamics will depend on the dimensionless synthetic magnetic flux, which takes real values modulo $2\pi$,
\begin{equation}\label{eq:path}
    \Phi_P = \sum_{\partial P} \phi_{ij} \equiv \frac{1}{\Phi_0}\iint_P \bm{B}\cdot d\bm{P},
\end{equation}
where the summation is taken along the closed oriented path $\partial P$ forming the boundary of a surface $P$, emulating a charged particle with magnetic flux quantum $\Phi_0$ in a magnetic field $\bm{B}$.
To demonstrate this, we consider four qubits arranged in a $2\times 2$ plaquette, labeled~A--D as shown in Fig.~\ref{fig:2x2}.
D is modulated at the detuning between D and qubits~B and~C (B and C are set to identical frequencies); B and~C are modulated at the detuning between themselves and~A.
The flux through the plaquette is the relative phase difference between the modulations of~B and~C, and is invariant to the modulation phase of qubit~D (which, according to Eq.~(\ref{eq:path}), contributes no net flux).
All other qubits in the array are deactivated by detuning them far ($\geq\SI{400}{MHz}$) from the active qubits.

We first consider the case $\Phi_P=0$.
After a particle is initialized at~A by a microwave $\pi$-pulse, it propagates to~D through a quantum walk~\cite{yan2019, gong2021, karamlou2022} along two trajectories, one via~B and the other via~C. As no relative phase is accumulated, the two trajectories constructively interfere, and the particle arrives at~D (Fig.~\ref{fig:2x2}, upper inset).
When the modulation of~B is shifted by $\pi$, inverting the sign of hopping between~B and~A so that $\Phi_P=\pi$, the two trajectories accumulate opposing signs and destructively interfere.
In this case, the particle cannot reach~D and reflects back to~A (Fig.~\ref{fig:2x2}, lower inset).
This effect is known as Aharonov-Bohm caging, and was recently demonstrated in a superconducting processor using different methodology~\cite{martinez2023}.

\begin{figure}[ht!]
\subfloat{\label{fig:chain}}
\subfloat{\label{fig:faraday}}
\includegraphics{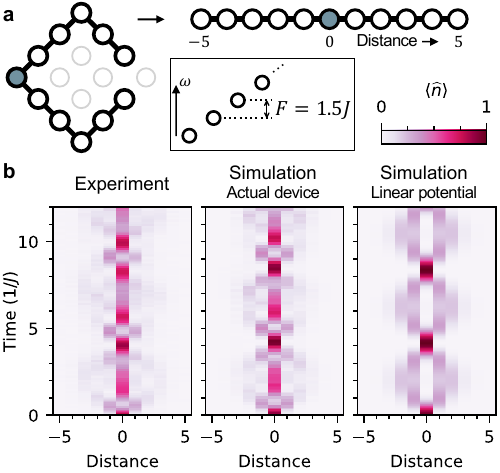}
\caption{\textbf{Faraday's law.} \textbf{(a)} A one-dimensional chain is formed by 11 sites along the periphery of the array. A particle is initialized at the center site (teal circle; distance $0$). Varying the modulation phases linearly in time at rate $1.5J$ generates a synthetic electric field $F$ throughout the chain, as if (inset) a linear potential is added to the on-site energies. \textbf{(b)} In the field, the particle undergoes Bloch oscillations and remains confined near the central site rather than propagating freely throughout the chain. Experimental data are compared to simulations of the actual experiment and of a tight-binding model with a linear potential.} 
\label{fig:fig4}
\end{figure}

%+++++++++++++++++++ 2x2, 3x3, 4x4 ++++++++++++++++++
Here, we can extend the experiment to arbitrary Peierls phases.
In Fig.~\ref{fig:2x2}, we present the population of~D as the modulation phase of~B is varied from $-\pi$ to $\pi$, yielding an Aharonov-Bohm interference pattern dependent on the synthetic magnetic flux~\cite{Aharonov1959}.
The parametric driving scheme can also be extended to larger rings.
In Fig.~\ref{fig:3x3} and Fig.~\ref{fig:4x4}, we present Aharonov-Bohm interference patterns in $8$- and $12$-site rings, respectively, with the enclosed qubits inactive (far detuned).
As before, these measurements begin with a single particle initialized on one corner of the ring, and we present the population on the opposing corner as a function of time and flux.
Again, we observe constructive interference at $\Phi_P=0$, destructive interference at $\Phi_P=\pm \pi$, and an intricate interference pattern between.

In Fig.~2 and throughout the remainder of this work, two layers of numerical simulations accompany experimental data.
We first present a decoherence-free simulation of our system as it is operated (``actual device"), using the lab-frame Hamiltonian Eq.~(\ref{eq:lab}) with measured values of each nearest- and next-nearest-neighboring bare exchange interaction strength $J_0^{ij}$.
For computational efficiency, each site is modeled as a two-level system.
The agreement between experimental and simulated results verifies faithful realization of the parametric coupling scheme on the superconducting processor.
Second, we present simulation of the idealized HH model Eq.~(\ref{eq:HH}) with uniform nearest-neighbor coupling only.
The qualitative similarity between experimental data and the latter simulation verifies that, with the parametric coupling scheme, we indeed emulate particles moving in a magnetic field.
Further simulations, presented in Section~S7 of the \supp{}, reveal that both the non-stationary terms in the rotating-frame Hamiltonian (those omitted from Eq.~(\ref{eq:rot})) and disorder in the exchange coupling strengths contribute to differences between dynamics of the device and of the idealized model.

\begin{figure*}[ht!]
\subfloat{\label{fig:hall_setup}}
\subfloat{\label{fig:hall_data}}
\subfloat{\label{fig:hall_coeff}}
\includegraphics{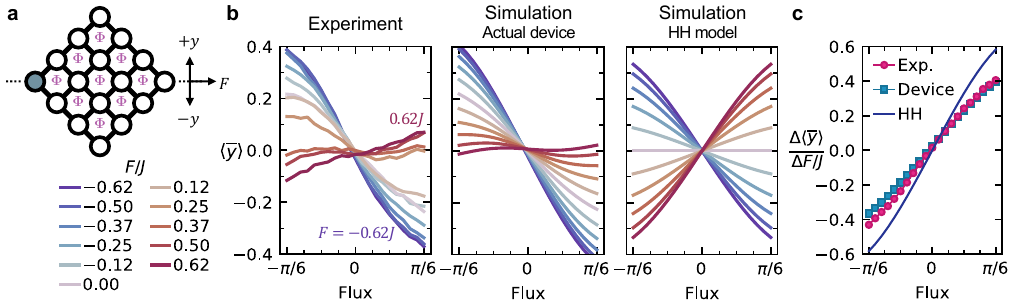}
\caption{\textbf{Hall effect in uniform synthetic magnetic and electric fields}. \textbf{(a)} One particle is initialized at a corner of the array (teal circle) in the presence of a synthetic longitudinal electric field $F$ in the $\hat x$~direction and a perpendicular ($\hat z$-axis) synthetic magnetic field $B_z$ created by threading flux $\Phi$ per plaquette. The system initially has a mirror symmetry about $y=0$ (dashed line). \textbf{(b)} Expected distance of the particle from the $y=0$ axis, time-averaged throughout the first $t=0.25/J$ of evolution after particle initialization. Data is shown as a function of synthetic flux per unit cell at various synthetic electric fields. Experimental data are accompanied by the simulated operation of the device and simulations of the idealized HH model plus a linear potential. \textbf{(c)} Hall coefficients, qualitatively analogous to Hall resistances, are extracted from the data in \textbf{(b)} by linear fits of $\langle \bar y \rangle$ versus synthetic electric field. Hall coefficients are shown as a function of synthetic magnetic flux.}
\label{fig:fig5}
\end{figure*}
%+++++++++++++++++++ gauge inv. +++++++++++++++++

As implied by Eq.~(\ref{eq:path}), the dynamics of our system should be invariant to the choice of Peierls phases provided that the flux $\Phi_P$ through each plaquette is conserved.
This property is a manifestation of the gauge invariance of magnetic vector potentials.
To verify gauge invariance,  we perform further Aharonov-Bohm interferometry experiments in the $8$-site ring, but with different arrangements of the Peierls phases.
In Fig.~\ref{fig:phase7}, the Peierls phase is placed on an exchange coupling on the upper trajectory across the ring, rather than the lower trajectory.
In Fig.~\ref{fig:phase3qb}, $\Phi_P$ is distributed between Peierls phases placed on three exchange couplings, each with value $\phi = \Phi_P/3$.
Both cases are related to the original case presented in Fig.~\ref{fig:3x3} by gauge transformations, i.e., by transformations $\bm{A}\rightarrow \bm{A}+\nabla \Lambda$ for scalar fields $\Lambda$ (depicted visually in the \supp{}).
The transformations conserve $\Phi_P$.
In all cases, we observe similar interference patterns, demonstrating the synthetic magnetic field upholds gauge freedom.

%+++++++++++++++++++ faraday ++++++++++++++++++

Up to this point, we have discussed the dynamics of particles when the synthetic vector potential is static.
A time-varying vector potential contributes to the electric field $\bm{E}$ according to Faraday's law of induction
\begin{equation}
    \bm{E} = -\frac{d\bm{A}}{dt}.
\end{equation}
Varying the Peierls phases in time as $\phi_i = \varphi t$, or, equivalently, shifting the modulation frequencies by $\varphi$, should therefore generate a synthetic electric field $F=-\varphi$ per site.
To test Faraday's law, we use 11~perimeter sites of the superconducting processor as a one-dimensional chain with open boundaries (Fig.~\ref{fig:chain}), and shift the modulation frequency by $\varphi = 1.5J$ for every exchange coupling.
We initialize a particle on the central site of the chain; its subsequent motion is displayed in Fig.~\ref{fig:faraday}.
Instead of propagating to the boundary of the chain and reflecting, the particle remains confined near the central site.
The confinement reflects Wannier-Stark localization of the particle from the synthetic electric field~\cite{preiss2015, guo2021, karamlou2022}, which can be understood in a momentum-space picture as a consequence of Bloch oscillations or in a real-space picture as confinement of the particle to sites having an effective energy within $\sim J$ of the initial site's energy.
Indeed, the localization dynamics match simulations of a tight binding model with a linear potential,
\begin{equation}\label{eq:F}
    \hat{H}_F/\hbar = \sum_{j=-5}^5 Fj\hat{n}_j+J\sum_{\langle i, j \rangle} \left(\hat{a}_i^\dag \hat{a}_j+\hat{a}_i \hat{a}_j^\dag \right), %\frac{\hat{H}_J}{\hbar},
\end{equation}
verifying that a synthetic electric field is established.
In Section~S7 of the \supp{}, we show that the somewhat stronger localization observed in the experiment, as compared to dynamics of the ideal model, is predominantly a consequence of non-stationary terms in the rotating-frame Hamiltonian.

%+++++++++++++++++++ Hall ++++++++++++++++++

The Hall effect is a well-known feature of electrical conductors in a perpendicular magnetic field where particles deflect transverse to the direction of current flow.
To emulate the Hall effect, we study the dynamics of a particle in the full $16$-site array with equal flux threading each plaquette, creating a uniform synthetic magnetic field $B_z$ (Fig.~\ref{fig:hall_setup}).
We additionally induce a synthetic longitudinal electric field $F$ per lattice constant via Faraday's law.
We initialize a particle at a corner of the lattice so that its initial average velocity ${\bm v}=v\hat x$ is longitudinal and preserves the mirror symmetry of the lattice about the line $y=0$ (dashed line in Fig.~\ref{fig:hall_setup}), and we measure the population of all sites as a function of time.

For an electrical conductor, the Hall voltage $V_y$ quantifies the transverse deflection of charge carriers in the presence of a magnetic field.
Here, we directly measure the average distance $\langle y \rangle$ of the particle from the $y=0$ axis, and consider its time-averaged value $\langle \bar y \rangle$ throughout the first $t = \SI{20}{ns} \approx 0.25/J$ of the experiment---approximately the time needed for the particle to reach the opposite corner of the array.
Data and corresponding simulations are shown at various magnetic and electric field strengths in Fig.~\ref{fig:hall_data}.

We first describe simulations of an idealized model comprising the HH model and a linear potential $F$.
When $F=0$, the particle is not deflected, even in finite magnetic field.
This feature is surprising: given the initial longitudinal velocity of the particle, from a classical description one would expect a transverse Lorentz force since $\bm{v}\times \bm{B}\neq 0$.
% In Section~S8 of the \supp{} we show through a semi-classical momentum-state expansion that symmetries of the particle's wavepacket imply that its net transverse velocity is odd in both $F$ and $B_z$.
% Therefore, even as the particle moves longitudinally across the lattice, only when $F$ and $B$ are both nonzero does the particle deflect transversely.
In Section~S8 of the \supp{}, we show through a semi-classical momentum-state expansion that this is not the case for a ballistic particle in a lattice.
Rather, $F$ is needed to break spatial inversion symmetry, $B_z$ is needed to break time-reversal symmetry, and only when both symmetries are broken does the particle deflect transversely.
The ensuing net transverse velocity is odd in both $F$ and $B_z$.

In the experiment, we instead observe transverse deflection when $F=0$. 
This difference is primarily due to inhomogeneity in $J_0^{ij}$ (see Section~S7 of the \supp{}), which breaks spatial inversion symmetry, and therefore skews the data as if the value of $F$ were offset.
Since coupling-strength inhomogeneity does not break time-reversal symmetry, it does not offset the value of $B_z$.
Aside from this skew, the deflection trends roughly linearly in flux and in $F$, matching the idealized model.
Simulations of the actual device closely reproduce the experimental results. 

To summarize these data, we consider the change in the transverse deflection $\Delta \langle \bar{y}\rangle$ per unit electric field $\Delta F$. 
In Fig.~\ref{fig:hall_coeff}, we present the linearized value of $\Delta \langle \bar{y}\rangle/\Delta F$ about the $F=0$ bias point, determined by the slope in a linear fit to the data in Fig.~\ref{fig:hall_data} at each value of synthetic magnetic flux (fiting is detailed in Section~S6 of the \supp{}).
This linearized quantity may be compared to, for example, a standard lock-in measurement of an electrical conductor's differential Hall resistance $R_{xy}=\Delta V_y/\Delta I_x$, which in general is proportional to magnetic field for small fields.
Surprisingly, in our system the transverse deflection trends with $F$ but not with the average longitudinal velocity, which is the same for $\pm F$ (data in Fig.~S20 of the \supp{}).
This behavior differs from dissipative electrical conductors where, according to the Drude model, the longitudinal drift velocity is proportional to electric field (so the transverse deflection trends with both).
The Hall effect in our system 
%is therefore inconsistent with a classical Lorentz force, which is often invoked to explain the Hall effect in conductors, but is consistent with a semi-classical description.
therefore cannot be described through the classical Lorentz force from the particle's net velocity.
The semi-classical description, which considers the evolution of the particle's wavepacket through the lattice's Brillouin zone, correctly predicts the observed behavior.

\section{Conclusion}

In this work, we emulate a tight-binding lattice in a spatially and temporally adjustable magnetic vector potential using a 2D array of superconducting qubits.
We verify the presence of the synthetic vector potential by observing Aharonov-Bohm interference in rings of various sizes, where transport through the two interferometer arms accumulates phases differing by the enclosed magnetic flux.
We confirm that the synthetic vector potential behaves consistently with two of Maxwell's equations: Gauss's law for magnetism $\nabla \cdot {\bm B}=0$, which is equivalent to a statement of its gauge freedom, and Faraday's law, which connects electric and magnetic fields.
Finally, we demonstrate transverse deflection of a particle traveling in the synthetic electromagnetic field.
% The transverse motion of the particle is a surprising manifestation of the Hall effect, as it does not match the traditional description of the Hall effect in terms of a classical Lorentz force.
While a classical Lorentz force is often invoked to explain the Hall effect, it does not accurately describe the dynamics of our system. 
Our results therefore showcase that the ballistic propagation of a quantum particle in a lattice differs from the motion of a free particle.
The realization of other members of the family of Hall effects, including the integer and fractional quantum Hall effects, is possible using the methods described in this work.

In the present experiment, we generate synthetic electromagnetic fields by parametrically modulating qubits that are coupled by fixed mutual capacitances.
The scheme may be viewed as the analog limit of a scheme recently used to approximate a magnetic field through repeated digital gates~\cite{neill2021}.
Alternatively, synthetic fields may be generated by modulating tunable-coupling elements~\cite{niskanen2006, chen2014, yan2018} with the qubits held static~\cite{roushan2017chiral}.
Tunable couplers confer simplified control of each parametrically-induced exchange coupling at the expense of additional hardware overhead.

The techniques introduced here extend beyond the emulation of natural 2D materials.
The synthetic magnetic flux through each unit cell of the qubit lattice may be individually set to any desired value modulo $2\pi$.
In contrast, creating magnetic-field patterns at the nanoscale or even mesoscale is difficult in natural materials, as is reaching high values of flux per unit cell.
For example, a half-magnetic-flux-quantum per unit cell in graphene would require approximately a \SI{22}{kT} field.
Our methodology therefore provides a new platform for creating artificial matter in high magnetic fields---for example, Hofstadter subband states~\cite{hofstadter1976, herzog2020, saito2021}---or complex magnetic environments---for example, the Haldane model~\cite{haldane1988, jotzu2014} and domain walls in Chern insulators~\cite{rosen2017, yasuda2017}.

{\em Note}: While preparing this manuscript, we became aware of related work using parametrically modulated coupling elements to generate a synthetic magnetic field~\cite{wang2024}.

\section{Acknowledgements}

The authors are grateful to Patrick M. Harrington, Jeffrey M. Gertler, Aaron L. Sharpe, Brice Bakkali-Hassani, Steven M. Girvin, Leonid S. Levitov, Xiao-Gang Wen, and Terry P. Orlando for fruitful discussions.
This material is based upon work supported in part by the U.S. Department of Energy, Office of Science, National Quantum Information Science Research Centers, Quantum System Accelerator (QSA); in part by the Defense Advanced Research Projects Agency under the Quantum Benchmarking contract; in part by U.S. Army Research Office Grant W911NF-23-1-0045; in part by the U.S. Department of Energy, Office of Science, National Quantum Information Science Research Centers, Co-design Center for Quantum Advantage (C2QA) under contract number DE-SC0012704; and in part by the Department of Energy and Under Secretary of Defense for Research and Engineering under Air Force Contract No. FA8702-15-D-0001.
ITR and MH are supported by an appointment to the Intelligence Community Postdoctoral Research Fellowship Program at the Massachusetts Institute of Technology administered by Oak Ridge Institute for Science and Education (ORISE) through an interagency agreement between the U.S. Department of Energy and the Office of the Director of National Intelligence (ODNI).
SM is supported by a NASA Space Technology Research Fellowship.
AC is partially supported by NSF DMR-2022428.
DAR acknowledges support from the National Science Foundation under award DMR-1747426.
Any opinions, findings, conclusions, or recommendations expressed in this material are those of the author(s) and do not necessarily reflect the views of the Department of Energy, the Department of Defense, or the Under Secretary of Defense for Research and Engineering.

\section{Author Contribution}

ITR, AC, MH, and MAD developed the theory for this work.
ITR, SM, and CNB performed the experiments.
AHK developed infrastructure with support from DAR.
RD, DKK, BMN, and MS fabricated the device with coordination from KS, MES, and JLY.
JAG and WDO provided technical oversight and support.
ITR wrote the manuscript with contributions from all authors.

\section{Competing Interests}

The Authors declare no competing interests.

\bibliography{refs}

\end{document}

\title{Supplementary Information for\\``Implementing a Synthetic Magnetic Vector Potential\\in a 2D Superconducting Qubit Array''}

\def\RLEaffil{Research Laboratory of Electronics, Massachusetts Institute of Technology, Cambridge, MA 02139, USA}
\def\LLaffil{MIT Lincoln Laboratory, Lexington, MA 02421, USA}
\def\Physaffil{Department of Physics, Massachusetts Institute of Technology, Cambridge, MA 02139, USA}
\def\EECSaffil{Department of Electrical Engineering and Computer Science, Massachusetts Institute of Technology, Cambridge, MA 02139, USA}

\author{Ilan~T.~Rosen}
\affiliation{\RLEaffil}

\author{Sarah~Muschinske}
\affiliation{\RLEaffil}
\affiliation{\EECSaffil}

\author{Cora~N.~Barrett}
\affiliation{\RLEaffil} 
\affiliation{\Physaffil}

\author{Arkya Chatterjee}
\affiliation{\Physaffil}

\author{Max~Hays}
\affiliation{\RLEaffil}

\author{Michael~DeMarco}
\affiliation{\Physaffil}

\author{Amir~Karamlou}
\affiliation{\RLEaffil}
\affiliation{\EECSaffil}

\author{David~Rower}
\affiliation{\RLEaffil}
\affiliation{\EECSaffil}

\author{Rabindra Das}
\affiliation{\LLaffil}

\author{David~K.~Kim}
\affiliation{\LLaffil}

\author{Bethany M. Niedzielski}
\affiliation{\LLaffil}

\author{Meghan~Schuldt}
\affiliation{\LLaffil}

\author{Kyle~Serniak}
\affiliation{\RLEaffil}
\affiliation{\LLaffil}

\author{Mollie E. Schwartz}
\affiliation{\LLaffil}

\author{Jonilyn L. Yoder}
\affiliation{\LLaffil}

\author{Jeffrey~A.~Grover}
\affiliation{\RLEaffil}

\author{William~D.~Oliver}
\affiliation{\RLEaffil}
\affiliation{\EECSaffil}
\affiliation{\Physaffil}

\maketitle

% \clearpage

\tableofcontents
\clearpage
\section{Parametric coupling scheme}

\subsection{Nonreciprocal coupling emulates a magnetic field}

Consider the Hamiltonian of a free charged particle in a magnetic field
\begin{equation}\label{eq:freeparticle}
    \hat{H}_\mathrm{free} = \frac{1}{2m}(\hat{\bm{p}}-q\bm{A})^2,
\end{equation}
where $m$ is mass, $\hat{\bm{p}}=\hbar \hat{\bm{k}}$ is the momentum operator, $\hat{\bm{k}}$ is the wavenumber operator, $q$ is charge, and $\bm{A}$ is the vector potential.
We wish to represent Eq.~(\ref{eq:freeparticle}) in the form of a tight-binding model on a lattice:
\begin{equation}
    \hat{H}_\mathrm{lattice}/\hbar = \sum_{\langle i,j\rangle} t_{ij}|\bm{r}_i\rangle\langle \bm{r}_j| + \mathrm{H.C.},
\end{equation}
where here we write $|\bm{r}_i\rangle$ to represent the position state corresponding to lattice site $i$.
For a square lattice with nearest-neighbor coupling and lattice spacing $a$, we can rewrite the Hamiltonian as:
\begin{equation}
    \hat{H}_\mathrm{lattice}/\hbar = \sum_i t_{i, i+x} |\bm{r}_i\rangle\langle \bm{r}_i+a\bm x| + t_{i, i+y} |\bm{r}_i\rangle\langle \bm{r}_i+a\bm y| + \mathrm{H.C.},
\end{equation}
where we have separately written the couplings in the $\bm x$ and $\bm y$ directions.
Using translation operators, we expand the position states:
\begin{equation}
    |\bm{r}_i\rangle\langle \bm{r}_i+a\bm{d}|= e^{i\hat{\bm{k}}\cdot a\bm{d}}|\bm{r}_i\rangle\langle \bm{r}_i|\approx \Big(1+ia\bm{\hat{k}}\cdot\bm{d}-\frac{1}{2}a^2(\bm{\hat{k}}\cdot\bm{d})^2\Big)|\bm{r}_i\rangle\langle \bm{r}_i|,
\end{equation}
where we consider to second order in $a$. If we take $t_{ij}=te^{-i\phi_{ij}}$ for real and uniform $t$, then:
\begin{align}
    & t_{i, i+x} |\bm{r}_i\rangle\langle \bm{r}_i+a\bm x| + t_{i, i+y} |\bm{r}_i\rangle\langle \bm{r}_i+a\bm y| \nonumber \\
    \approx\ & t\left[(1-i\phi_{i,i+x}-\frac{\phi_{i,i+x}^2}{2})(1+i\hat{k}_xa-\frac{1}{2}\hat{k}_x^2a^2)+(1-i\phi_{i,i+y}-\frac{\phi_{i,i+y}^2}{2})(1+i\hat{k}_ya-\frac{1}{2}\hat{k}_y^2a^2)\right]|\bm{r}_i\rangle\langle \bm{r}_i| \nonumber\\
    \approx\ & t \left(2+i(\hat{k}_xa+\hat{k}_ya-\phi_{i,i+x}-\phi_{i,i+y} )    - \frac{1}{2}(\phi_{i,i+x}^2+\phi_{i,i+y}^2+\hat{k}^2a^2) +\phi_{i,i+x}\hat{k}_xa +\phi_{i,i+y}\hat{k}_ya      \right)|\bm{r}_i\rangle\langle \bm{r}_i|
\end{align}

We note that $\phi_{ij}$ in general has position dependence that is not periodic in translations by $a$, so $\phi_{ij}$ may not commute with the momentum operator (just as $\bm{A}$ may not commute with $\hat{\bm{p}}$). Ignoring a constant term, and defining $\bm{\phi}_i = (\phi_{i, i+x}, \phi_{i, i+y})$, the Hamiltonian is then:
\begin{equation}
    \hat{H}_\mathrm{lattice} = -\hbar t\sum_{i} (\phi_i^2+\hat{k}^2a^2 - \bm{\phi}_{i}\cdot\hat{\bm{k}}a - \hat{\bm{k}}\cdot\bm{\phi}_{i}a)|\bm{r}_i\rangle\langle \bm{r}_i| = -\frac{ta^2}{\hbar}\sum_{i} (\hat{\bm{p}}-\frac{\hbar}{a}\bm{\phi}_{i})^2|\bm{r}_i\rangle\langle \bm{r}_i|.
\end{equation}
Therefore, $\hat{H}_\mathrm{lattice}$ is the lattice representation of $\hat{H}_\mathrm{free}$ under the identifications
\begin{align}
    t &= -\frac{\hbar }{ 2a^2m}\\
    \phi_{ij} &= \frac{1}{\hbar}\int_{\bm{r}_i}^{\bm{r}_j} q\bm{A}\cdot d\bm{r},
\end{align}
where $i$ and $j$ are neighboring sites and $r_{i,j}$ are their positions. Note that the Peierls phases $\phi_{ij}$ are agnostic to $q$ and $\bm{A}$ individually, but depend only upon their product. It is convenient to define the magnetic flux quantum as $\Phi_0 = \hbar/q$. Then the magnetic flux through a surface $P$,
\begin{equation}
    \Phi_B(P) = \oint_{\partial P} \bm{A}\cdot d\bm{r},
\end{equation}
where $\partial P$ is the boundary of $P$, can be expressed as a dimensionless quantity $\Phi(P) = \frac{\Phi_B(P)}{\Phi_0}$. Mapping the dimensionless flux onto the lattice model yields:
\begin{equation}
    \Phi(P) \rightarrow \Phi_P = \sum_{\partial P} \phi_{ij},
\end{equation}
so that:
\begin{equation}\label{eq:phip}
    \Phi_P = \frac{1}{\Phi_0} \oint_{\partial P} \bm{A}\cdot d\bm{r}.
\end{equation}

\subsection{Parametric modulation induces nonreciprocal coupling}\label{sec:coupling_derivation}

Here, we derive the appearance of Peierls phases form the parametric coupling scheme. Consider two two-level systems, with destruction operators $\hat{a}$ and $\hat{b}$, that are transversely coupled with strength $J_0$. If site $\hat{a}$ is sinusoidally modulated with frequency $\zeta$, amplitude $\Omega$, and phase $\phi$, the Hamiltonian is
\begin{equation}
    \hat{H}/\hbar = \big(\omega_a - \Omega\cos(\zeta t +\phi) \big) \hat{a}^\dagger \hat{a} + \omega_b \hat{b}^\dagger \hat{b} + J_0(\hat{a}^\dagger \hat{b} + \hat{b}^\dagger \hat{a}).
\end{equation}
Now transform to the instantaneous rotating frame of the qubits using the unitary transformation
\begin{equation}
    U = \exp\left[i \left(\omega_a t - \frac{\Omega}{\zeta}\sin(\zeta t +\phi) \right) \hat{a}^\dagger \hat{a} + i \omega_b t \hat{b}^\dagger \hat{b} \right].
\end{equation}
The Hamiltonian in this rotating frame is
\begin{equation}
    \hat{H}_R = U\hat{H}U^\dagger + i\hbar \dot U U^\dagger = \hbar J_0(\tilde a^\dagger \tilde b + \tilde b^\dagger \tilde a),
\end{equation}
where $\tilde a = U\hat{a}U^\dagger$ and $\tilde b = U\hat{b}U^\dagger$. It is straightforward to show that $e^{i\alpha \hat{a}^\dagger \hat{a}}\hat{a} e^{-i\alpha \hat{a}^\dagger \hat{a}} = e^{-i\alpha} \hat{a}$ for some constant $\alpha$. Therefore:
\begin{align}
    \tilde a &= e^{i\frac{\Omega}{\zeta}\sin(\zeta t+\phi) - i\omega_a t}\hat{a} \nonumber\\
    \tilde b &= e^{-i\omega_b t}\hat{b}.
\end{align}
Defining $\delta = \omega_a - \omega_b$:
\begin{equation}
    \hat{H}_R/\hbar = J_0\left(e^{i\delta t}e^{-i\frac{\Omega}{\zeta}\sin(\zeta t+\phi)}\hat{a}^\dagger \hat{b} + e^{-i\delta t}e^{i\frac{\Omega}{\zeta}\sin(\zeta t+\phi)}\hat{a} \hat{b}^\dagger \right).
\end{equation}

To simplify, we use the Jacobi-Anger identity
\begin{equation}
    e^{iz \sin\theta} = \sum_{n=-\infty}^\infty \mathcal{J}_n(z)e^{in\theta},
\end{equation}
where $\mathcal{J}_n(z)$ is the $n$th Bessel function of the first kind, arriving at
\begin{equation}\label{eq:full_HR}
    \hat{H}_R /\hbar= J_0 e^{i\delta t} \sum_{n=-\infty}^\infty \mathcal{J}_n\left(\frac{\Omega}{\zeta}\right) e^{-in(\zeta t+\phi)}\hat{a}^\dagger \hat{b} + \textrm{H.C.}.
\end{equation}

Finally, we choose to set the modulation frequency equal to the detuning between the two sites, i.e. $\zeta=\delta$. This choice selects the $n=1$ term in the Jacobi-Anger expansion to be stationary (other terms in the expansion may be made stationary by choosing $\delta$ to be integer multiples of $\zeta$). The stationary component of the rotating-frame Hamiltonian is then
\begin{equation}
    \hat{H}_S/\hbar = J_0 \mathcal{J}_1\left(\frac{\Omega}{\delta}\right) e^{-i\phi}\hat{a}^\dagger \hat{b} + \textrm{H.C.},
\end{equation}
while other terms rotate at multiples of $\delta$. This shows that the parametric modulations induce transverse coupling at an effective rate
\begin{equation}\label{eq:effective_rate_1tone}
    J(\Omega) = J_0 \mathcal{J}_1\left(\frac{\Omega}{\delta}\right)
\end{equation}
and with a Peierls phase $\phi$.

\clearpage
\subsection{Parametric modulation shifts qubit frequency}\label{sec:shifts}

The stationary component of a transmon qubit's frequency shifts when the qubit is modulated. We show here that this is a consequence of the nonlinearity in the transmon's Hamiltonian.

The Hamiltonian of an asymmetric tunable transmon threaded by a sinusoidally-varying flux $\varphi$ may be written as
\begin{equation}
\hat{H} = 4E_C \hat{n}^2 - \Bigg[E_{J\Sigma}\cos\Big(\frac{\pi \Phi_{\rm ext}}{\Phi_S}\Big)\sqrt{1 + d^2 \tan^2\Big(\frac{\pi \Phi_{\rm ext}}{\Phi_S}\Big)}\Bigg]\cos(\hat{\phi} - \phi_0),
\end{equation}
where $\hat n$ and $\hat \phi$ are the charge and flux operators, respectively, $E_C$ is the charging energy, $E_{J\Sigma}$ is the total Josephson energy, $\Phi_{\rm ext}$ is the external flux threading the SQUID loop, $\Phi_S=h/2e$ is the superconducting flux quantum, $d$ is the junction asymmetry parameter, and $\tan(\phi_0) = d\tan(\pi\Phi_{\rm ext}/\Phi_S)$~\cite{koch2007}. Let $\varphi = 2\pi \Phi_{\rm ext} / \Phi_S$ and consider sinusoidal flux modulation $\varphi = \varphi_0 +\Theta\cos(\delta t)$. Note that we here represent the amplitude of the modulation by $\Theta$ rather than $\Omega$ to emphasize that we here write the amplitude in units of dimensionless flux.

Without loss of generality, we can transform the basis as $\hat \phi \rightarrow \hat\phi - \phi_\mathrm{offset}$ where $\phi_\mathrm{offset}$ is given by $\tan(\phi_\mathrm{offset}) = d\tan(\varphi_0/2)$.
Together with the condition $\Theta\ll 2\pi$, this transform enables the small-angle approximation:
\begin{equation}
\hat{H} = 4E_C \hat{n}^2 - \Bigg[E_{J\Sigma}\cos\Big(\frac{\varphi}{2}\Big)\sqrt{1 + d^2 \tan^2\Big(\frac{\varphi_0}{2}\Big)}\Bigg]\cos(\hat{\phi} - \frac{d}{2}\Theta\cos(\delta t)).
\end{equation}
Expanding:
\begin{equation}
\hat{H} = 4E_C \hat{n}^2 - E_{J\Sigma}(\Theta)\left[\cos(\hat{\phi})\cos\left( \frac{d}{2}\Theta\cos(\delta t)\right)+\sin(\hat{\phi})\sin\left( \frac{d}{2}\Theta\cos(\delta t)\right) \right],
\end{equation}
where
\begin{equation}
    E_{J\Sigma}(\Theta) \approx E_{J\Sigma} \left[ \cos\left(\frac{\varphi_0}{2}\right)\cos\left(\frac{\Theta}{2}\cos(\delta t)\right) + \sin\left(\frac{\varphi_0}{2}\right)\sin\left(\frac{\Theta}{2}\cos(\delta t)\right) \right]\sqrt{1 + d^2 \tan^2(\frac{\varphi_0}{2})}
\end{equation}
Using the Jacobi-Anger expansions
\begin{align}
    \cos(a\cos b) &= \mathcal{J}_0(a) + 2\sum_{m=1}^\infty(-1)^m\mathcal{J}_{2m}(a)\cos(2mb)\nonumber\\
    \sin(a\cos b) &= -2\sum_{m=1}^\infty(-1)^m\mathcal{J}_{2m-1}(a)\cos\big((2m-1)b\big),
\end{align}
where $\mathcal{J}_n$ is the $n$th Bessel function of the first kind, we can then write the stationary component of the qubit Hamiltonian:
\begin{equation}
    \hat{H}_{\rm DC} = 4E_C \hat{n}^2 -\Bigg[E_{J\Sigma}\mathcal{J}_0\left(\frac{\Theta}{2}\right)\cos\left(\frac{\varphi_0}{2}\right)\sqrt{1 + d^2 \tan^2\left(\frac{\varphi_0}{2}\right)}\Bigg]\mathcal{J}_0\left(d\frac{\Theta}{2}\right)\cos(\hat{\phi}).
\end{equation}
This shows that the modulation reduces the effective Josephson energy:
\begin{equation}
    \frac{E^\mathrm{eff}_{J\Sigma}}{E_{J\Sigma}} = \mathcal{J}_0\left(\frac{\Theta}{2}\right)\mathcal{J}_0\left(\frac{d\Theta}{2}\right)\leq 1,
\end{equation}
lowering the stationary component of the qubit's frequency. The shift in frequency goes as $\Theta^2$ to lowest order, since $\mathcal{J}_0(a) = 1-\frac{a^2}{4}+O(a^4)$ and the qubit frequency is $\omega=\left(\sqrt{8E^\mathrm{eff}_{J\Sigma}E_C}-E_C\right)/\hbar$.

\clearpage
\subsection{Spectroscopy of a parametrically modulated qubit}\label{sec:spec}

\begin{figure*}[ht!]
\includegraphics{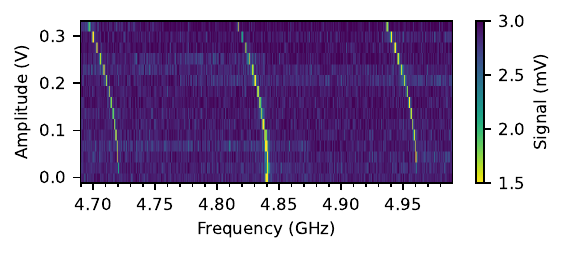}
\caption{\textbf{Two-tone spectroscopy of a modulated qubit}. The response signal to a probe tone applied to the resonator of qubit~13, while a second tone is applied to excite the qubit. The qubit is flux biased to a DC frequency setpoint $\SI{4.84}{GHz}$. The frequency of the second tone is swept while the qubit is sinusoidally modulated at \SI{120}{MHz} with varying amplitude.} 
\label{sfig:modspec_wide}
\end{figure*}

To visualize the effects discussed in the previous two sections, in Fig.~\ref{sfig:modspec_wide}, we present two-tone spectroscopy of a qubit as a function of the amplitude with which it is modulated. The modulation frequency here is $\zeta=\SI{120}{MHz}$. At zero modulation amplitude, the qubit responds to the spectroscopy tone when the frequency of the tone matches the qubit's DC frequency setpoint, here $\omega_0=\SI{4.84}{GHz}$. When the modulation amplitude $\Omega$ is finite, the frequency at which the qubit responds shifts downwards, $\omega(\Omega)<\omega_0$, due to the effect discussed above. Additionally, the qubit responds at two more frequencies $\omega\pm \zeta$. The qubit may therefore be parametrically coupled to a second qubit by placing the second qubit on resonance with either sideband.

\subsection{Frequency layout and choice of gauge}\label{sec:layout}

\begin{figure*}[ht!]
\subfloat{\label{sfig:layout_numbering}}
\subfloat{\label{sfig:layout_modulation_frequencies}}
\includegraphics{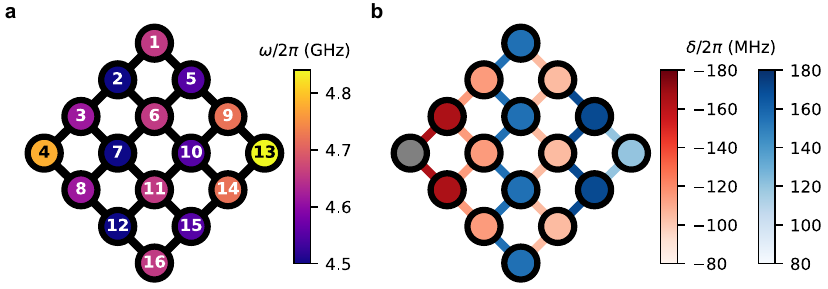}
\caption{\textbf{Frequency layout}. \textbf{(a)} A schematic of the $4\times 4$ transmon array, with each qubit represented by a circle. The orientation matches diagrams in the main text figures. Each qubit is indexed by a number between 1 and 16; these indices are used to refer to specific qubits throughout the \supp{}. The color of each qubit describes its DC frequency setpoint. \textbf{(b)} Each qubit is colored according to the frequency at which it is modulated, and each nearest-neighbor bond is colored according to the detuning between the two qubits. Negative detunings, moving from left to right, are colored in red shades (left colorbar), while positive detunings are colored in blue shades (right colorbar).} 
\label{sfig:layout}
\end{figure*}

Fig.~\ref{sfig:layout}a shows the DC frequency setpoints $\omega_i$ of all qubits. All qubits are operated at a frequency below their respective upper sweet spots by roughly $\delta_i$ or more, ensuring the modulation amplitudes may approach $\delta_i$ to provide large effective coupling rates. The signs of the detunings between neighboring qubits are staggered so that qubits may be placed near their upper sweet spots (still satisfying the prior condition), where dephasing rates are lower. Indices for all qubits, used throughout the \supp{} to reference individual qubits, are also shown in Fig.~\ref{sfig:layout}a. Note that in Fig.~1a of the main text, the bottom left qubit is qubit~1.

Fig.~\ref{sfig:layout}b shows the detunings between qubits and corresponding modulation frequencies $\delta_i$. The modulation on each qubit activates exchange coupling to nearest-neighboring qubits detuned by the same frequency; for example, the modulation on qubit~13 activates hopping to qubits~9 and 14. The color indicates whether the detuning is positive or negatively signed; for example, the frequency of qubit~13 is higher than the frequency of qubits~9 and 14. Qubit~4 is not modulated.

\begin{figure*}[ht!]
\includegraphics{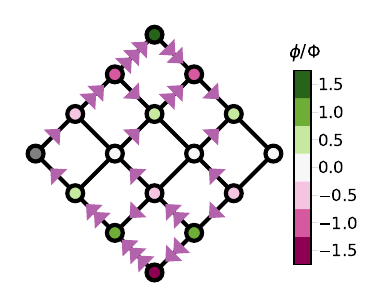}
\caption{\textbf{Modulation phase layout and choice of gauge}. To realize a uniform synthetic magnetic field with $\Phi$ flux per plaquette, the modulation tone for each qubit is given the phase $\phi_i$ indicated by the qubit's color; for example, the modulation tone sent to qubit~16 has a phase $-1.5\Phi$. The modulation phases generate the Peierls phases indicated by the purple arrows on each bond, where each arrow represents a Peierls phase $\phi = \Phi/2$. The flux in each plaquette is the oriented sum of the Peierls phases around the plaquette; to define the sign of the flux, we take the sum along the counterclockwise oriented path.} 
\label{sfig:layout_phases}
\end{figure*}

Fig.~\ref{sfig:layout_phases} describes the pattern of modulation phases used to create a uniform synthetic magnetic field throughout the lattice (used for the experiment presented in Fig.~5 of the main text). Choosing the pattern of modulation phases sets the gauge of the magnetic vector potential. We note that the sign of the Peierls phase depends on the sign of the modulation phase as well as on the sign of the detuning between the two qubits.

There are many possible choices of the frequency layout and the gauge. The layouts used in the present experiment were chosen for convenience given the parameters of the device and the desire to keep the DC operation frequencies of the qubits high to reduce dephasing rates.

\clearpage
\subsection{Table of equivalences}

To summarize the parametric coupling scheme, in Table~\ref{tab:equiv} we present physical quantities alongside their equivalent in our experiment.

\begin{table}[h]
\begin{center}
    \begin{tabular}{|c|c|}\hline
     & \\
    Physical quantity & Synthetic quantity\\
    & \\ \hline
    & \\
    Particle & Qubit excitation \\
    & \\
    Lattice constant $a$ & Dimensionless \\
    & \\
    $\frac{q}{\hbar}\int_{\bm{r}_i}^{\bm{r}_j} \bm{A}\cdot d\bm{r}$ & $\phi_{ij}$ \\
    & \\
    $\frac{1}{\Phi_0} \oint_{\partial P} \bm{A}\cdot d\bm{r}$ & $\Phi_P = \sum_{\partial P} \phi_{ij}$\\
    & \\
    $qa^2B/\hbar$* & $\Phi_P$ \\
    & \\
    $\frac{q}{\hbar}\int_{\bm{r}_i}^{\bm{r}_j} \frac{d\bm{A}}{dt}\cdot d\bm{r}$ & $\frac{d\phi_{ij}}{dt}$ \\
    & \\
    $qEa$* & $F$ \\
    & \\ \hline
    \end{tabular}
    \caption{\textbf{Table of equivalences} between physical values and their synthetic counterpart. Note that in our scheme, the charge $q$ of the particle being emulated is not chosen, rather the product of the charge and the electromagnetic field is set. *For uniform fields.}
    \label{tab:equiv}
\end{center}
\end{table}

\clearpage
\section{Device parameters}

\subsection{Measurement setup}

\begin{figure*}[ht!]
\includegraphics[width = 0.85\textwidth]{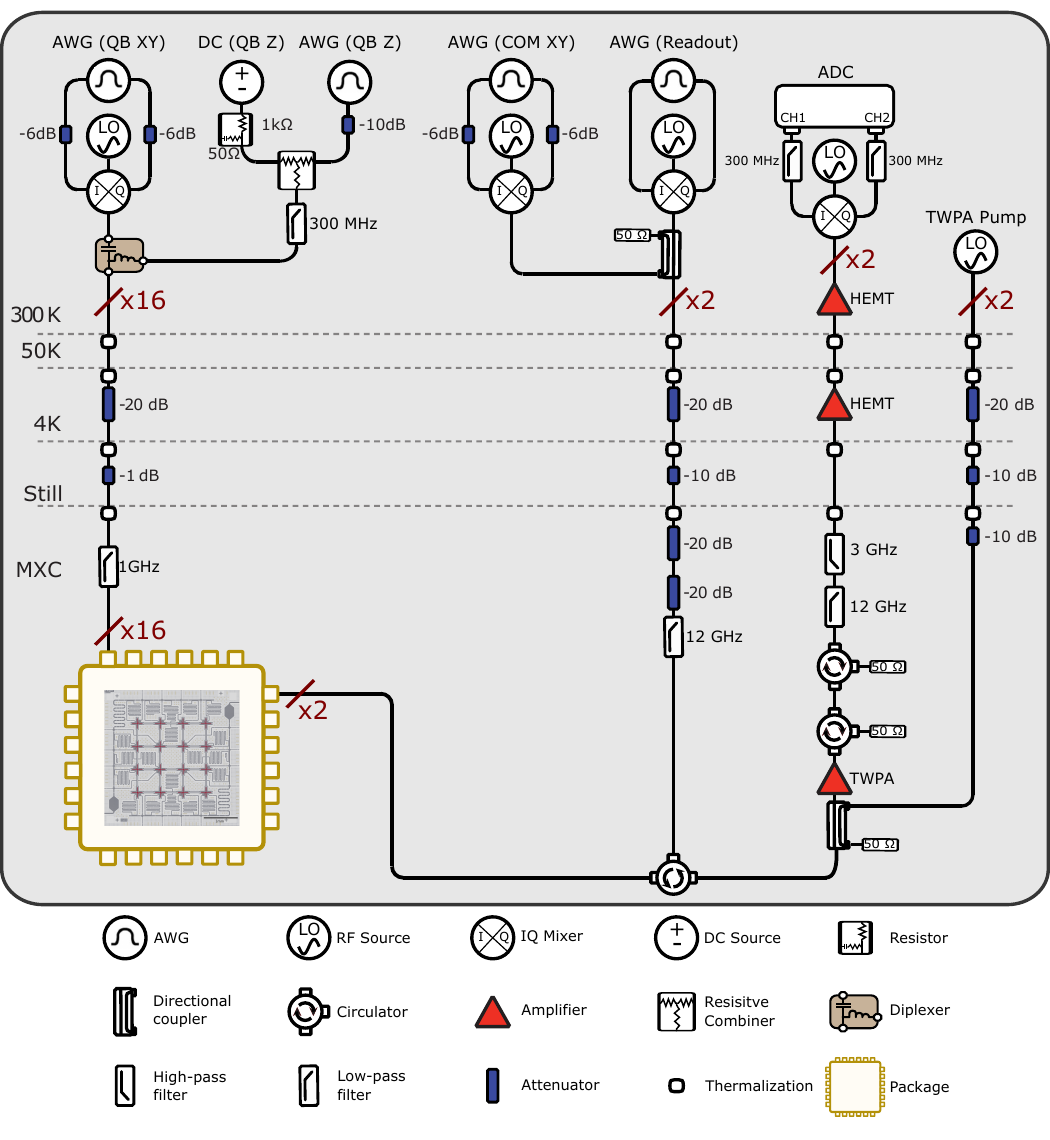}
\caption{\textbf{Diagram of the measurement setup} (adapted from Ref.~\cite{karamlou2024}). } 
\label{sfig:wiring}
\end{figure*}

Measurements are conducted using a BlueFors XLD-600 dilution refrigerator with a base temperature of approximately $\SI{22}{mK}$. The flip-chip-geometry device is housed in a 24-port copper microwave package with a PCB interposer and SMP connectorization. Signal and ground traces are connected to the interposer tier of the device via aluminum wirebonds. A diagram of the measurement electronics and wiring is presented in Fig.~\ref{sfig:wiring}; further details are provided in Ref.~\cite{karamlou2024}. A summary of device parameters is presented in Table~\ref{tab:sample_parameters}.

\subsection{Bare coupling rates}

\begin{figure*}[ht!]
\includegraphics{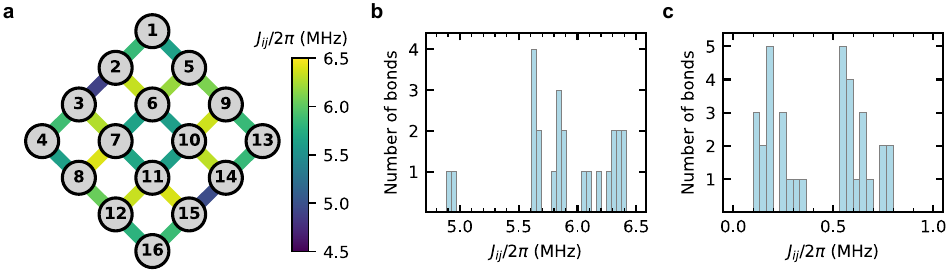}
\caption{\textbf{Bare coupling rates}. \textbf{(a)} Coupling rates $J_0^{ij}/2\pi$ between nearest-neighbor qubits. Couplings between next-nearest-neighbor qubits are not shown. \textbf{(b)} Histogram of bare coupling rates between nearest neighbors (adapted from Ref.~\cite{karamlou2024}). \textbf{(c)} Histogram of bare coupling rates between next-nearest neighbors.} 
\label{sfig:bare_J}
\end{figure*}

The bare exchange coupling rates $J_0^{ij}$ are determined by the fixed mutual capacitance between pairs of qubits. Experimentally measured values of the bare coupling rates are shown in Fig.~\ref{sfig:bare_J}. The nearest-neighbor couplings have average strength $\SI{5.9}{MHz}$ with standard deviation $\SI{0.4}{MHz}$. The next-nearest-neighbor couplings have average strength $\SI{0.43}{MHz}$ with standard deviation $\SI{0.23}{MHz}$.

\subsection{Coherence times}

\begin{figure*}[ht!]
\subfloat{\label{sfig:depolarization}}
\subfloat{\label{sfig:ramsey}}
\subfloat{\label{sfig:echo}}
\includegraphics{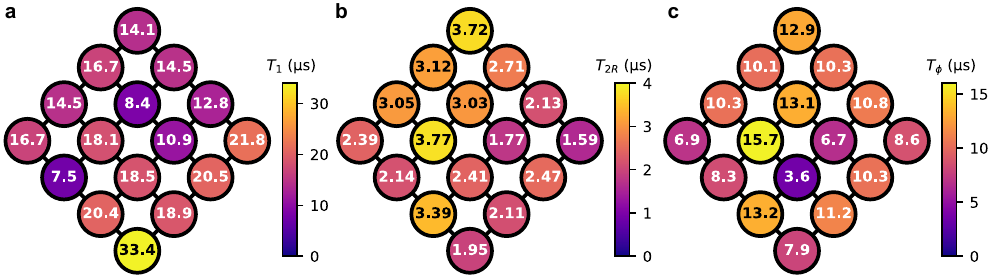}
\caption{\textbf{Decoherence times} for each qubit at its operation frequency. \textbf{(a)} Depolarization times. \textbf{(b)} Ramsey dephasing times. \textbf{(c)} Pure dephasing times, determined using Hahn echo sequences.} 
\label{sfig:decoherence}
\end{figure*}

The depolarization times of all qubits at the operational frequency setpoints is shown in Fig.~\ref{sfig:depolarization}, with a mean value of $\SI{16.7}{\mu s}$. The Ramsey dephasing times times of all qubits at the operational frequency setpoints is shown in Fig.~\ref{sfig:ramsey}, with a mean value of $\SI{2.61}{\mu s}$. The pure dephasing times of all qubits at the operational frequency setpoints is shown in Fig.~\ref{sfig:echo}, with a mean value of $\SI{10.0}{\mu s}$. 

% \begin{figure*}[ht!]
% \includegraphics{sfig/sfig_depolarization.pdf}
% \caption{\textbf{Depolarization times} for each qubit at its operation frequency.} 
% \label{sfig:depolarization}
% \end{figure*}

% \begin{figure*}[ht!]
% \subfloat{\label{sfig:ramsey}}
% \subfloat{\label{sfig:echo}}
% \includegraphics{sfig/sfig_decoherence.pdf}
% \caption{\textbf{Decoherence times} for each qubit at its operation frequency. \textbf{(a)} Ramsey dephasing times. \textbf{(b)} Pure dephasing times, determined using Hahn echo sequences.} 
% \label{sfig:decoherence}
% \end{figure*}

\begin{table}[ht!]
\centering
{\renewcommand{\arraystretch}{1.7}   %modifying vertical spacing in tabular
\begin{tabular}{ p{4.5cm}  p{1.5cm} p{1.5cm}  p{1.5cm}  p{1.5cm} p{1.5cm} p{1.5cm}  p{1.5cm}  p{1.5cm}  }
\toprule

%%%%%%%%%%%%%%%%
Parameters & QB1 & QB2 & QB3 & QB4 & QB5 & QB6 & QB7 & QB8 \\
\hline
Feedline & A & A & A & A & A & A & B & B \\
$\omega_{\mathrm{res}}/2\pi$ (GHz) & 6.206 & 6.358 & 6.239 & 6.497 & 6.365 & 6.339 & 6.428 & 6.281 \\
$\omega_{\mathrm{q}}^{\mathrm{max}}/2\pi$ (GHz) & 4.859 & 4.873 & 4.781 & 4.833 & 4.691 & 4.825 & 4.695 & 4.939 \\
$\omega_{\mathrm{q}}^{\mathrm{set}}/2\pi$ (GHz) & 4.655 & 4.500 & 4.615 & 4.780 & 4.550 & 4.655 & 4.500 & 4.615 \\
Mod. freq. $\delta/2\pi$ (MHz) & 155 & 115 & 165 & N.A. & 105 & 155 & 115 & 165 \\
Mod. phase $\phi$ & $1.5\Phi$ & $-\Phi$ & $-0.5\Phi$ & N.A. & $-\Phi$ & $0.5\Phi$ & 0 & $0.5\Phi$ \\
$T_1$ ($\SI{}{\micro s}$) & 14.1 & 16.7 & 14.5 & 16.7 & 14.5 & 8.4 & 18.1 & 7.5 \\
$T_{2R}$ ($\SI{}{\micro s}$) & 3.7 & 3.1 & 3.1 & 2.4 & 2.7 & 3.0 & 3.7 & 2.1 \\
$T_\phi$ ($\SI{}{\micro s}$) & 12.9 & 10.1 & 10.3 & 6.9 & 10.2 & 13.1 & 15.7 & 8.3 \\
$\mathcal{F}_{\mathrm{gg}}$ & 0.97 & 0.96 & 0.96 & 0.97 & 0.95 & 0.96 & 0.89 & 0.96 \\
$\mathcal{F}_{\mathrm{ee}}$ & 0.93 & 0.93 & 0.93 & 0.92 & 0.91 & 0.93 & 0.81 & 0.91 \\
\hline
\hline
Parameters & QB9 & QB10 & QB11 & QB12 & QB13 & QB14 & QB15 & QB16 \\
\hline
Feedline & A & A & B & B & B & B & B & B \\
$\omega_{\mathrm{res}}/2\pi$ (GHz) & 6.282 & 6.429 & 6.338 & 6.357 & 6.502 & 6.250 & 6.361 & 6.197 \\
$\omega_{\mathrm{q}}^{\mathrm{max}}/2\pi$ (GHz) & 5.065 & 4.967 & 4.894 & 4.838 & 5.074 & 5.008 & 4.771 & 4.947 \\
$\omega_{\mathrm{q}}^{\mathrm{set}}/2\pi$ (GHz) & 4.720 & 4.550 & 4.655 & 4.500 & 4.840 & 4.720 & 4.550 & 4.655 \\
Mod. freq. $\delta/2\pi$ (MHz) & 170 & 105 & 155 & 115 & 120 & 170 & 105 & 155 \\
Mod. phase $\phi$ & $0.5\Phi$ & $0$ & $-0.5\Phi$ & $\Phi$ & $0$ & $-0.5\Phi$ & $\Phi$ & $-1.5\Phi$ \\
$T_1$ ($\SI{}{\micro s}$) & 12.8 & 10.9 & 18.5 & 20.4 & 21.8 & 20.5 & 18.9 & 33.4 \\
$T_{2R}$ ($\SI{}{\micro s}$) & 2.1 & 1.8 & 2.4 & 3.4 & 1.6 & 2.5 & 2.1 & 1.9 \\
$T_\phi$ ($\SI{}{\micro s}$) & 10.8 & 6.7 & 3.6 & 13.2 & 8.6 & 10.3 & 11.2 & 7.9 \\
$\mathcal{F}_{\mathrm{gg}}$ & 0.95 & 0.96 & 0.97 & 0.93 & 0.97 & 0.97 & 0.96 & 0.97 \\
$\mathcal{F}_{\mathrm{ee}}$ & 0.70 & 0.93 & 0.93 & 0.91 & 0.93 & 0.92 & 0.91 & 0.91 \\
\hline
\hline
\end{tabular}
}
\caption{\textbf{Summary of measurement parameters and device performance}. Performance metrics are measured at the frequency setpoints used in the present experiment. The following are shown: 
to which of the two readout resonator feedlines each qubit is linked, 
the readout resonator frequencies $\omega_{\mathrm{res}}$, 
the maximum transmon transition frequencies $\omega_{\mathrm{q}}^{\rm max}$ at the upper flux-insensitive point, 
the transmon transition frequency setpoints for the experiment $\omega_{\mathrm{q}}^{\rm set}$, 
the modulation frequency $\delta$ applied to each qubit for parametric coupling,
the corresponding modulation phase $\phi$ applied to thread a uniform flux $\Phi$ through each plaquette,
the measured qubit decay times $T_1$, Ramsey coherence times $T_{2R}$s, dephasing times $T_\phi$ extracted from Hahn echo sequences, 
and the measurement fidelities $\mathcal{F}_{gg}$ ($\mathcal{F}_{ee}$) of measuring the qubit the ground (excited) state after preparing it in the ground (excited) state. 
Readout fidelities are primarily limited by state preparation error due to thermal qubit population; $\mathcal{F}_{ee}$ is additionally limited by decay during readout.}
\label{tab:sample_parameters}
\end{table}
%%%%%%%%%%%%%%%%%

\clearpage
\section{Tuneup procedure}

Procedures for calibrating the flux crosstalk matrix, corrections for flux pulse transients, pulse alignment, and readout are discussed in Refs.~\cite{barrett2023, karamlou2024}. The parametric modulation tones are generated by arbitrary wave generators (AWGs) at room temperature, and are sent to the qubits through qubit-specific flux bias lines. We use a four-step process to calibrate the parametric modulation tones.

\begin{figure*}[ht!]
\subfloat{\label{fig:tuneup_modspec_1}}
\subfloat{\label{fig:tuneup_modspec_2}}
\includegraphics{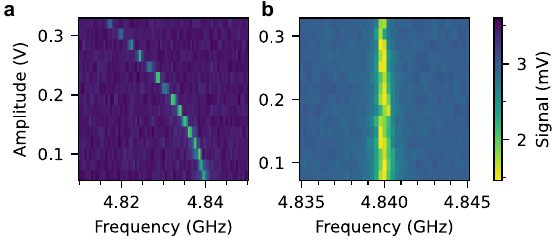}
\caption{\textbf{Frequency shift correction}. \textbf{(a)} Two-tone spectroscopy of qubit~13 as a function of modulation tone amplitude. The qubit is flux biased to a DC setpoint of $\SI{4.84}{GHz}$. \textbf{(b)} Two-tone spectroscopy of qubit~13 while, at each modulation amplitude, applying a correction to the qubit's DC frequency setpoint based on the measurement in \textbf{(a)}.} 
\label{sfig:tuneup_modspec}
\end{figure*}

\subsection{Frequency shift correction}

In Sections~\ref{sec:shifts} and~\ref{sec:spec}, we showed that the qubits' effective frequencies shift when they are modulated. To compensate, we adjust the DC flux biases applied to each qubit to return all qubits to their intended frequencies. To determine mappings between modulation amplitudes and frequency shifts, we perform two-tone spectroscopy of each qubit at a variety of modulation amplitudes, as shown in Fig.~\ref{fig:tuneup_modspec_1} for qubit~13. Each qubit's effective frequency is determined at each amplitude by a Lorentzian fit to the measured response as a function of probe tone frequency. DC corrections are determined by quadratic fits to the effective frequency as a function of amplitude. These DC corrections are applied in all subsequent measurements.

\subsection{Frequency shift correction -- second round} 

A second round of correction to each qubit's DC frequency setpoint is determined by repeating the prior measurement while applying the DC corrections determined from the first measurement (Fig.~\ref{fig:tuneup_modspec_2}). A second level of DC corrections are determined by quadratic fits to the effective frequency as a function of amplitude. Both stages of DC corrections are applied in all subsequent measurements.

\begin{figure*}[ht!]
\subfloat{\label{sfig:tuneup_hopping_9}}
\subfloat{\label{sfig:tuneup_hopping_14}}
\subfloat{\label{sfig:tuneup_hopping_fit}}
\includegraphics{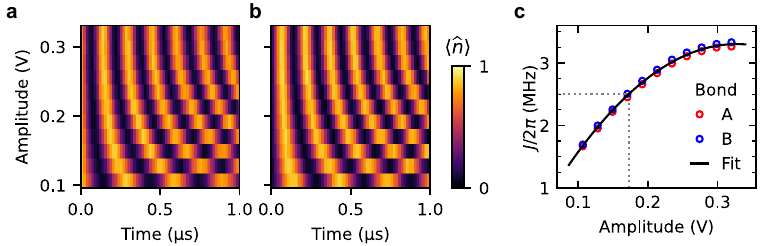}
\caption{\textbf{Hopping rate calibration}. \textbf{(a)} Population of qubit~9 following a $\pi$-pulse on qubit~9 as a function of evolution time and modulation amplitude of qubit~13, with all other qubits far detuned. \textbf{(b)} Population of qubit~14 following a $\pi$-pulse on qubit~13 as a function of evolution time and modulation amplitude of qubit~13, with all other qubits far detuned. \textbf{(c)} The hopping rate at each modulation amplitude is extracted by sinusoidal fitting to population versus time. Bond A (red circles) represents hopping between qubits~13 and 9; bond B (blue circles) represents hopping between qubits~13 and 14; solid line is a quadratic fit. The horizontal dashed line denotes the target hopping rate, and the vertical dashed line indicates the modulation amplitude chosen to provide the target hopping rate, based on the quadratic fit.} 
\label{sfig:tuneup_hopping}
\end{figure*}

\subsection{Determining parametric modulation amplitude}\label{sec:tuneup_amp}

As described in Section~\ref{sec:layout}, for each bond between nearest-neighboring qubits, exchange interactions are parametrically induced by modulating one of the two qubits. To determine the modulation amplitude appropriate to generate coupling at the desired rate, a particle is initialized at one of the two sites. The probability of the particle being on each of the two qubits is measured as a function of evolution time at various modulation amplitudes, with the remaining 14~qubits inactive (far detuned). For qubits whose modulation parametrically induces coupling across two bonds (see Fig.~\ref{sfig:layout_modulation_frequencies}), this process is repeated independently for both bonds (Fig.~\ref{sfig:tuneup_hopping_9},~\ref{sfig:tuneup_hopping_14}).

During the evolution time, the particle swaps between the two qubits. A sinusoidal fit to the population of one qubit as a function of time yields the hopping rate for each modulation amplitude. A quadratic fit to the hopping rate versus amplitude is then used to choose the amplitude needed to provide the desired hopping rate. For qubits whose modulation activates two bonds, a quadratic fit to the average hopping rate between the two bonds at each amplitude is used (Fig.~\ref{sfig:tuneup_hopping_fit}).

\begin{figure*}[ht!]
\includegraphics{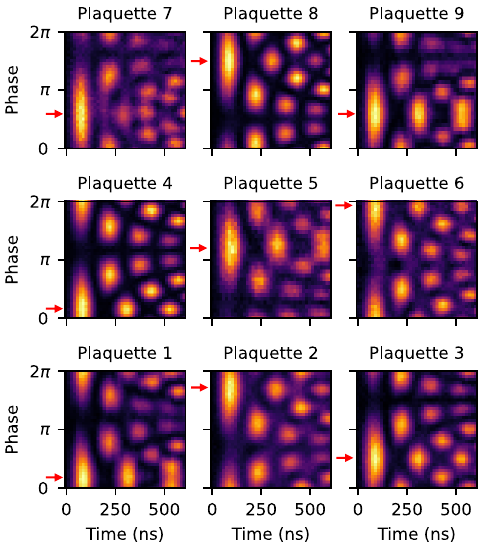}
\caption{\textbf{Phase offset calibration}. Aharonov-Bohm interference patterns are shown for all nine plaquettes in the qubit array. Here, the vertical axis represents the phase of the modulation of one qubit as set by room-temperature control electronics, with zero phase for the other qubits in the plaquette. Constructive interference occurs when the phase corrects for the signal delay lengths between control electronics and the qubits, and is indicated by the red arrows. Destructive interference occurs at $\pi$ out of phase.} 
\label{sfig:tuneup_plaquettes}
\end{figure*}

\subsection{Parametric modulation phase calibration} 

The flux through each plaquette is determined by the relative phase of two qubits that are modulated at the same frequency (see Section~\ref{sec:layout}). Because signal delay lengths may not be identical, the relative phases of the AWG outputs will, in general, be offset from the relative phase at which the tones arrive at the qubits. To measure this offset, we measure the interference pattern of a particle moving in each plaquette as a function of the phase of one qubit, as in Fig.~2a of the main text. The population of the qubit opposite the initial location of the particle is shown for each plaquette as a function of phase and time in Fig.~\ref{sfig:tuneup_plaquettes}. The phase at which the particle fully reaches this qubit represents constructive interference---zero relative phase between the two modulation tones. In subsequent experiments, these offset phases are added as corrections to the desired Peierls phases to ensure that all plaquettes are threaded by the desired synthetic flux.

\clearpage
\section{Effective hopping rates and hopping rate disorder}\label{sec:effective_j}

\begin{figure*}[ht!]
\includegraphics{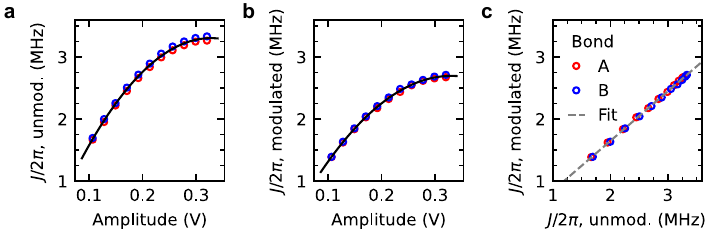}
\caption{\textbf{Reduction in the hopping rate when both qubits are modulated.} Experimentally determined hopping rates induced by parametrically modulating qubit~13, extracted as described in Fig.~\ref{sfig:tuneup_hopping}. Bond A (red circles) represents hopping between qubits~13 and 9, while bond B (blue circles) represents hopping between qubits~13 and 14. For measurements of each bond, the third qubit is far detuned. \textbf{(a)} Hopping rates when qubits~9 and 14 are not modulated as a function of modulation amplitude. \textbf{(b)} Hopping rates when qubits~9 and 14 are modulated at a different frequency than the modulation of qubit~13, as indicated in Fig.~\ref{sfig:layout}, as a function of amplitude. Solid lines in \textbf{(a,~b)} are quadratic fits. \textbf{(c)} Comparison of the hopping rates with and without the second qubits modulated at each modulation amplitude. A linear fit (dashed line) indicates the hopping rate is reduced by a factor of $1.25$ when the second qubit is modulated.} 
\label{sfig:hopping_with_modulation}
\end{figure*}

In Section~\ref{sec:tuneup_amp}, we described characterizing the hopping rate of each nearest-neighbor bond as a function of the amplitude of the modulation tone inducing parametric coupling across the bond. This characterization was done with all other modulation tones off, a choice made to simplify tuneup. However, when the second qubit is also modulated (to induce coupling across another bond; see Fig.~\ref{sfig:layout}), the coupling rate across the first bond decreases. For example, we characterized the coupling rate between qubits~13 and 9 when only qubit~13 is modulated. Yet in a full experiment, qubit~9 is also modulated to induce coupling between qubits~9 and 5. When qubit~9 is modulated, the hopping rate between qubits~13 and 9 decreases.

This effect may be intuited as a consequence of the second qubit's spectral weight being transferred from its central frequency to sidebands as the modulation amplitude increases (see Fig.~\ref{sfig:modspec_wide}; the central frequency of the second qubit aligns with a sideband of the first qubit generated by its modulation), therefore, the effective coupling between the two qubits decreases. To demonstrate this effect, in Fig.~\ref{sfig:hopping_with_modulation} we experimentally determine the hopping rates of the bonds activated by modulation of qubit~13 with and without the second qubits (qubits~9 and 14) modulated (at the amplitude determined in the aforementioned calibration procedure). We observe a reduction in the effective hopping rate by a factor of $1.25$.

The reduction of the effective hopping rate by a factor of $1.25$ may also be understood analytically. We choose the modulation amplitude $\Omega_i$ for each qubit $i$ to provide coupling to qubit $j$ with strength $\SI{2.5}{MHz}$ when all other qubits are stationary. By Eq.(~\ref{eq:effective_rate_1tone}), this sets $\mathcal{J}_1\left(\frac{\Omega_i}{\delta_i}\right) = J(\Omega_i)/J_0 =0.42$, or $\frac{\Omega_i}{\delta_i}=0.94$. Similarly, we choose $\frac{\Omega_j}{\delta_j}=0.94$.

Now consider the experimental condition where qubits $i$ and $j$ are simultaneously modulated. Repeating the rotating frame transformation described in Section~\ref{sec:coupling_derivation} using the instantaneous rotating frame of both qubits yields the rotating frame Hamiltonian:
% \begin{equation}
%     H_R & = g e^{-i \delta t}e^{i\frac{\Omega_1}{\zeta_1}  \sin (\zeta_1 t +\phi_1) } e^{-i\frac{\Omega_2}{\zeta_2}  \sin (\zeta_2 t +\phi_2) } a_1^\dag a_2 +  \text{H.C.}
% \end{equation}
\begin{equation}\label{eq:hr_full_2modulated}
    H^{ij}_{R,\mathrm{full}} = J_0 e^{-i \delta_i t} \Bigg[ \sum_{n=-\infty}^{\infty} \mathcal{J}_n\bigg(\frac{\Omega_i}{\delta_i}\bigg) e^{in(\delta_i t+\phi_i)}\Bigg] \Bigg[ \sum_{m=-\infty}^{\infty} \mathcal{J}_m\bigg(\frac{\Omega_j}{\delta_j}\bigg) e^{-im(\delta_j t+\phi_j)}\Bigg]\hat{a}_i^\dag \hat{a}_j +  \text{H.C.}
\end{equation}
The stationary component, corresponding to the $n=1$ and $m=0$ term, is:
\begin{align}
    H_R^{ij} = J_0 \mathcal{J}_1\bigg(\frac{\Omega_i}{\zeta_i}\bigg) \mathcal{J}_0\bigg(\frac{\Omega_j}{\zeta_j}\bigg) e^{i\phi_i} \hat{a}_i^\dag \hat{a}_j +  \text{H.C.}
\end{align}
Therefore, the effective coupling strength is $J=J_0 \mathcal{J}_1\bigg(\frac{\Omega_i}{\zeta_i}\bigg) \mathcal{J}_0\bigg(\frac{\Omega_j}{\zeta_j}\bigg)$. Using the modulation amplitudes chosen above, the coupling strength is adjusted by a factor of $\mathcal{J}_0\bigg(\frac{\Omega_j}{\zeta_j}\bigg)=0.8$ (i.e., reduced by a factor of $1.25$), yielding the effective coupling strength $J=\SI{2.0}{MHz}$.

We note that Eq.~(\ref{eq:hr_full_2modulated}) includes terms rotating at $\delta_i-\delta_j$. To reduce the effect of these terms, we have chosen $|\delta_i-\delta_j|\gg J$ for adjacent qubits $i$ and $j$.

In all data presented in the main text, the modulation amplitudes were chosen to provide a hopping rate of $\SI{2.5}{MHz}$ via the aforementioned tuneup procedure (i.e. when the second qubit is not modulated). Therefore, the effective hopping rate (i.e. the hopping rate when the second qubit is modulated at a frequency that does not match the detuning between the two qubits) is approximately $\SI{2.0}{MHz}$ for all bonds except those involving qubit~4, as qubit~4 is not modulated.

\begin{figure*}[t]
\includegraphics{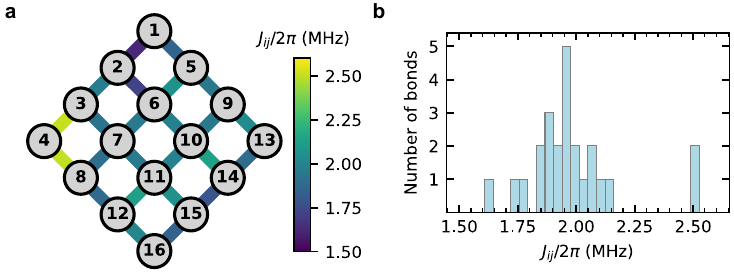}
\caption{\textbf{Simulated effective hopping rates throughout the lattice.} Here, each qubit is simulated as a two-level system. The modulation amplitude of each qubit is determined by simulating the same tuneup procedure as used in the experiment. Then the effective hopping rate across each bond is determined with all qubits modulated (except qubit~4, which is not modulated in the experiment). \textbf{(a)} Spatial visualization of the effective hopping rates throughout the lattice, with the color each bond reflecting the effective hopping rate between the two qubits it connects. \textbf{(b)} Histogram of the effective hopping rate of all 24~nearest-neighbor bonds. Note the effective hopping rates between qubit~4 and qubits~3 and~8 are higher than that of other bonds because qubit~4 is not modulated.} 
\label{sfig:effective_J}
\end{figure*}

To understand the effective hopping rate throughout the lattice, we present simulations of the effective hopping rate of all bonds in Fig.~\ref{sfig:effective_J}. For these simulations, the tuneup procedure described in Section~\ref{sec:tuneup_amp} is replicated, and the experimentally-determined bare coupling rate $J_0^{ij}$ (which is set by the physical geometry of the chip) of each bond is included. As each lattice site is represented in the simulation as a two-level system, compensation for DC frequency shifts are unneeded. Comparing the effective hopping rates to the bare hopping rates described in Fig.~\ref{sfig:bare_J} shows that the parametric coupling scheme partially mitigates coupling strength disorder (since the coupling rates are tuned by selecting modulation amplitudes), but not fully (since there are 24~bonds but only 15~modulation amplitudes). Ignoring bonds involving qubit~4, simulations indicate the average effective nearest-neighbor coupling strength has standard deviation $\SI{0.1}{MHz}$.

\clearpage
\section{Visualizing gauge transformations}

Gauge transforms of the continuous magnetic vector potential $A$ are transformations of the form
\begin{equation}
    \mathbf{A}\rightarrow \mathbf{A}+\nabla \Lambda
\end{equation}
for a scalar field $\Lambda$. The flux through each plaquette is invariant to gauge transformations, which can be shown using Eq.~(\ref{eq:phip}) and Stokes' theorem:
\begin{equation}
    \Phi_P \rightarrow \frac{1}{\Phi_0} \oint_{\partial P} (\mathbf{A}+\nabla \Lambda) \cdot d\mathbf{r} = \frac{1}{\Phi_0} \oint_{\partial P} \mathbf{A} \cdot d\mathbf{r} + \frac{1}{\Phi_0} \iint_{P} \nabla\times\nabla \Lambda \cdot d\mathbf{r} = \Phi_P.
\end{equation}

\begin{figure*}[ht!]
\includegraphics{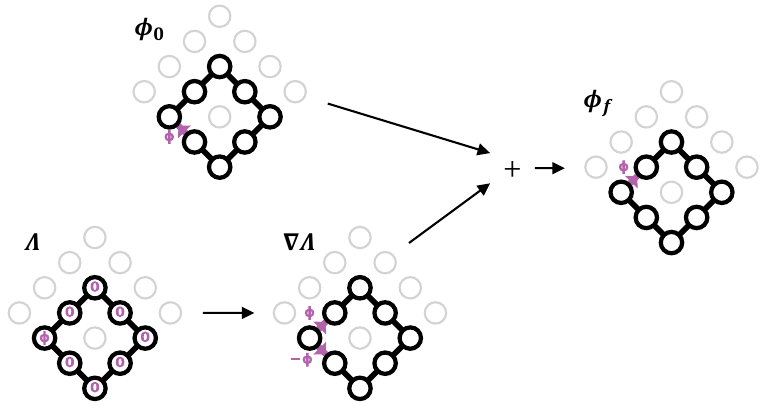}
\caption{\textbf{Visualizing the discretized magnetic gauge transform} used for Fig.~3a of the main text. The initial gauge $\phi_0=\{\phi_{ij}^0\}$ describes the Peierls phases used in Fig.~2b of the main text to thread a flux $\Phi_P=\phi$ through the $8$~site ring. Adding the gradient of a discretized scalar field $\Lambda$ transforms to the gauge $\phi_f$ used in Fig.~3a of the main text.} 
\label{sfig:gauge_transform_1}
\end{figure*}

\begin{figure*}[ht!]
\includegraphics{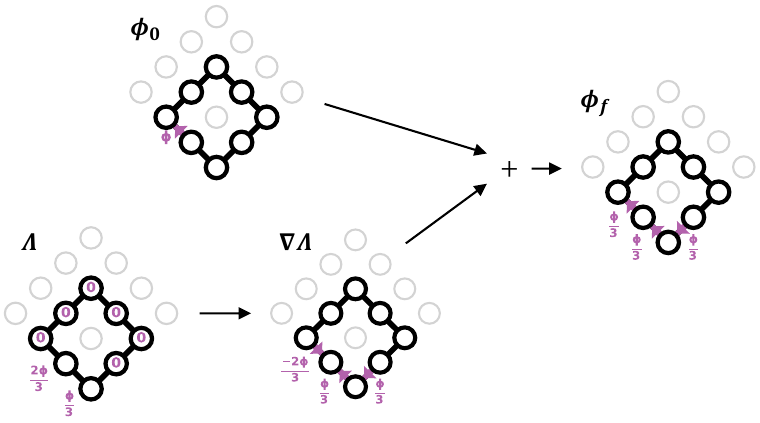}
\caption{\textbf{Visualizing the discretized magnetic gauge transform} from the gauge used in Fig.~2b of the main text to the gauge used in Fig.~3b of the main text.} 
\label{sfig:gauge_transform_2}
\end{figure*}

Under the gauge transformation given by the scalar field $\Lambda$, the Peierls phases transform as
\begin{equation}
    \phi_{ij} \rightarrow \phi_{ij} + \frac{1}{\Phi_0}\int_{r_i}^{r_j} \nabla \Lambda \cdot d\mathbf{r}.
\end{equation}
To visualize a gauge transformation on a lattice, we can discretize $\Lambda$ by considering its value at each lattice site $\Lambda_i$. Then the Peierls phases transform as
\begin{equation}
    \phi_{ij} \rightarrow \phi_{ij} + \frac{a}{\Phi_0}(\Lambda_j - \Lambda_i),
\end{equation}
where $a$ is the lattice constant. In Figs.~\ref{sfig:gauge_transform_1} and~\ref{sfig:gauge_transform_2}, we present visualizations of the gauge transformations used in Fig.~3 of the main text, for convenience working in units where $\Phi_0=a=1$.

\clearpage
\section{Extended data}

\begin{figure*}[ht!]
\includegraphics{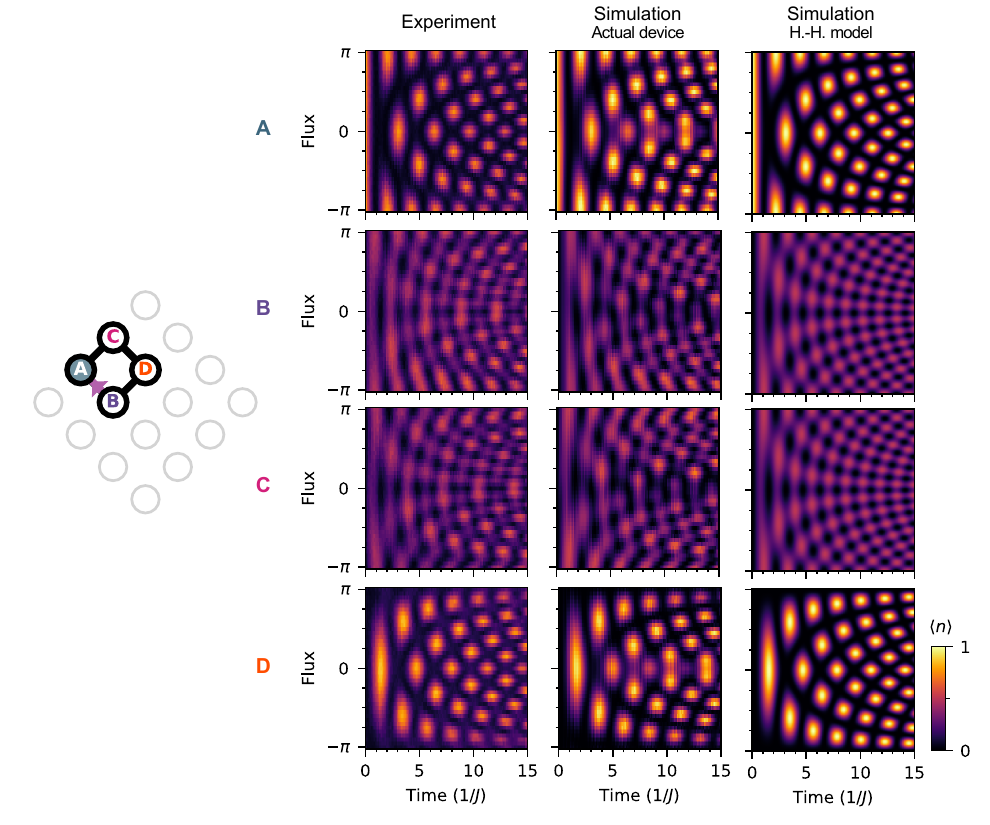}
\caption{\textbf{Interference in a $2\times 2$ plaquette---extended results}. The same dataset presented in Fig.~2a of the main text, but with the population of all four qubits shown. Experimental data are accompanied by simulations of the actual device and of the idealized Harper-Hofstadter model.} 
\label{sfig:extended_P2}
\end{figure*}

\begin{figure*}[ht!]
\includegraphics{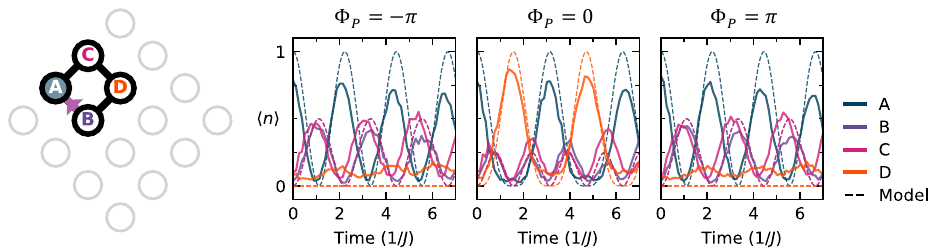}
\caption{\textbf{Interference in a $2\times 2$ plaquette}. Line cuts of the dataset presented in Fig.~2a of the main text at fluxes $-\pi$, $0$, and $\pi$. Experimental data (solid lines) are accompanied by simulations of the idealized Harper-Hofstadter model (dashed lines).} 
\label{sfig:cuts_P2}
\end{figure*}

\begin{figure*}[ht!]
\includegraphics{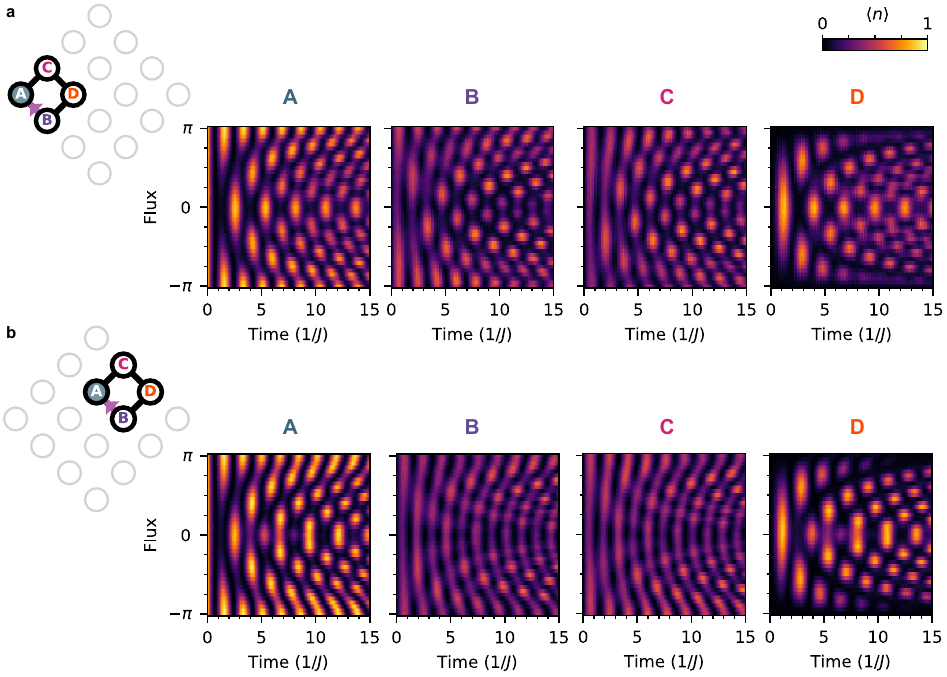}
\caption{\textbf{Interference in other $2\times 2$ plaquettes}. \textbf{(a)} Aharonov-Bohm interference as in Fig.~2a, but in the leftmost plaquette. \textbf{(b)} Aharonov-Bohm interference as in Fig.~2a, but in the upper right plaquette. Experimental data are presented for all four qubits in each plaquette.} 
\label{sfig:extended_P3_P4}
\end{figure*}

\begin{figure*}[ht!]
\includegraphics{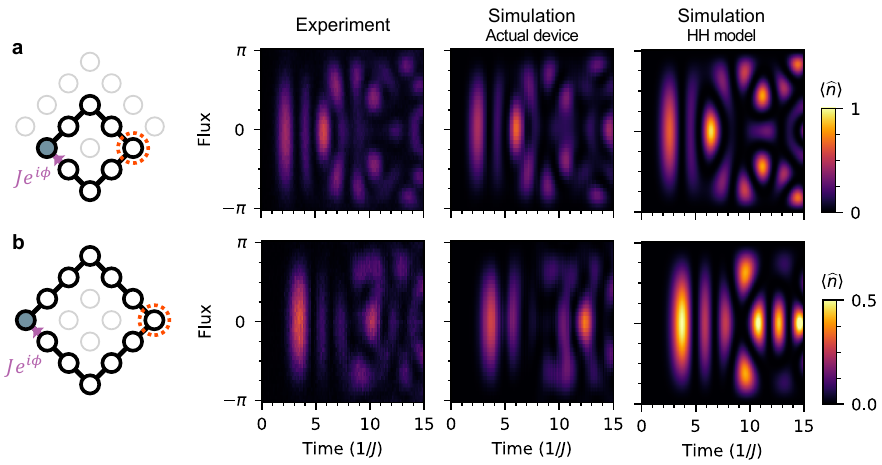}
\caption{\textbf{Interference in rings}. The data shown in Fig.~2(b, c) of the main text, shown with experimental and simulated results on the same color scales. \textbf{(a)} An $8$-site ring. \textbf{(b)} A $12$-site ring.} 
\label{sfig:AB_uniform_color}
\end{figure*}

The dataset demonstrating Aharonov-Bohm interference in a $2\times 2$ plaquette, presented in Fig.~2a of the main text, is reproduced in Fig.~\ref{sfig:extended_P2}, but here including the measured population of all four qubits in the plaquette. Figure~\ref{sfig:cuts_P2} displays line cuts of the data at zero and $\pi$ flux. Measurements of Aharonov-Bohm interference in two more $2\times 2$ plaquettes are presented in Fig.~\ref{sfig:extended_P3_P4}. The Aharonov-Bohm interferometry results in larger rings presented in Fig.~2b and Fig.~2c of the main text are reproduced in Fig.~\ref{sfig:AB_uniform_color} with experimental and simulated results displayed in uniform color scales.

\begin{figure*}[ht!]
\subfloat{\label{sfig:extended_3x3_gauge}}
\subfloat{\label{sfig:extended_3x3_gauge2}}
\subfloat{\label{sfig:extended_3x3_gauge3qb}}
\subfloat{\label{sfig:extended_3x3_cancel}}
\includegraphics{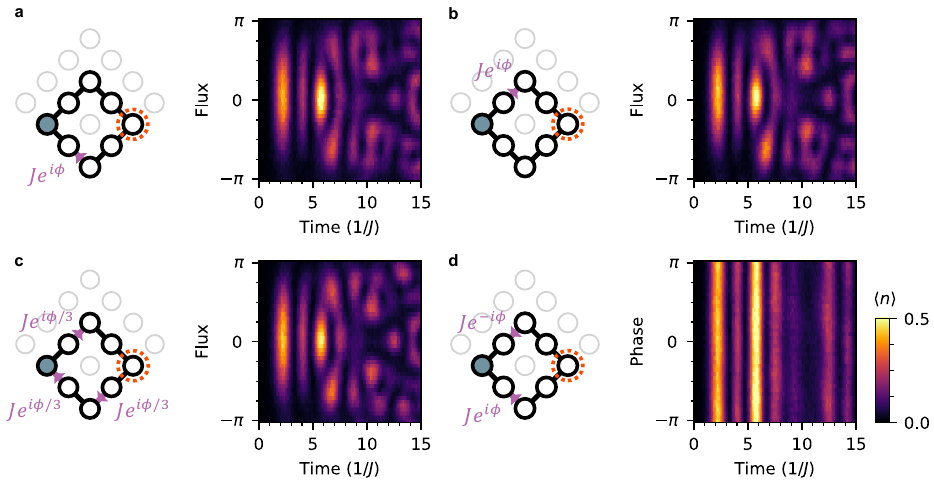}
\caption{\textbf{Interference in an $8$-site ring---extended results}. \textbf{(a, b)} Aharonov-Bohm interference as in Fig.~2b and Fig.~3 of the main text, with the Peierls phase placed on different bonds. \textbf{(c)} Interference when the flux is distributed as phases placed on three bonds in a different arrangment than Fig.~3b of the main text. \textbf{(d)} Interference when equal Peierls phases are placed on two bonds, but with opposite signs (opposing signs here defined with respect to an oriented path around the ring). As the phases vary, the net flux through the ring remains zero, so the interference pattern does not significantly change. Experimental data is shown in all panels.} 
\label{sfig:extended_3x3}
\end{figure*}

In Fig.~\ref{sfig:extended_3x3_gauge} and Fig.~\ref{sfig:extended_3x3_gauge2}, we present Aharonov-Bohm interference in the same $8$-site ring discussed in Fig.~2b and Fig.~3 of the main text, but under two more choices of gauge. In Fig.~\ref{sfig:extended_3x3_gauge3qb}, we show similar results as Fig.3b of the main text, but in a gauge where the Peierls phases are distributed between another choice of the three qubits. In Fig.~\ref{sfig:extended_3x3_cancel}, we present a measurement where two Peierls phases are varied such that the net flux remains zero. As expected, the interference does not change substantially as the phases are varied.

\begin{figure*}[ht!]
\includegraphics{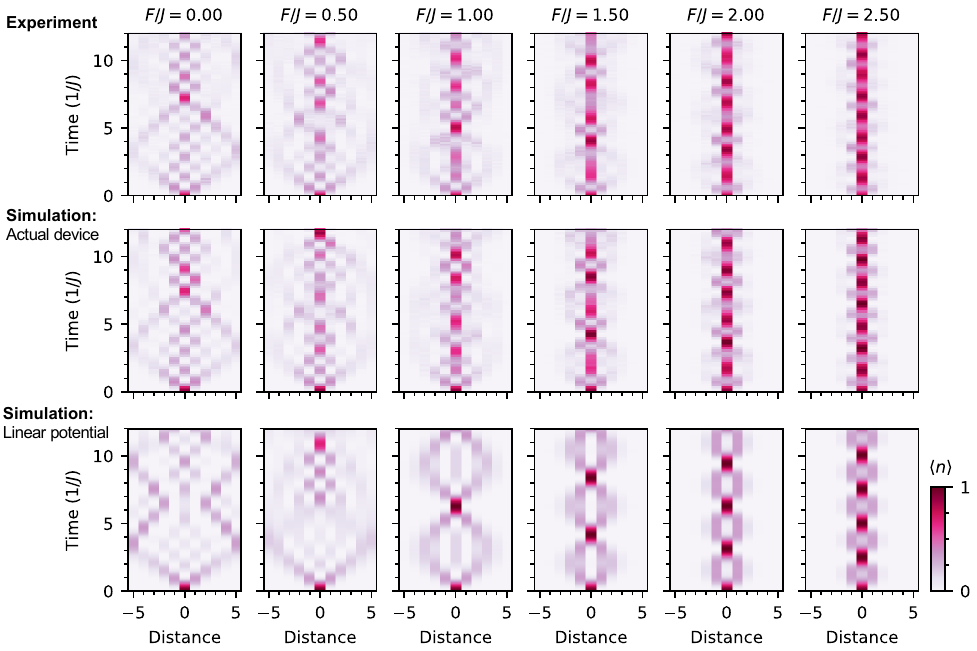}
\caption{\textbf{Localization at various synthetic electric field strengths}. Population at each site along an 11-site one-dimensional chain as a function of time after being initialized at the central site. Dynamics at various synthetic electric field strengths are shown, as indicated at top in units of the effective hopping rate $J$. Experimental data are accompanied by simulations of the actual device and of an idealized model with a linear potential.} 
\label{sfig:efield}
\end{figure*}

In Fig.~\ref{sfig:efield}, we present data from the experiment discussed in Fig.~4 of the main text---the dynamics of a one-dimensional chain under a synthetic electric field---at more values of the electric field. When the electric field is zero, the particle is itinerant. After being initialized at the central site, the particle propagates to the edges of the chain and reflects. As the electric field approaches $J$ per site, Wannier-Stark localization of the particle becomes apparent: the particle does not reach the edges of the chain. At higher electric fields, the localization length decreases and the particle's dynamics are tightly confined around its initial position.

\begin{figure*}[ht!]
\includegraphics{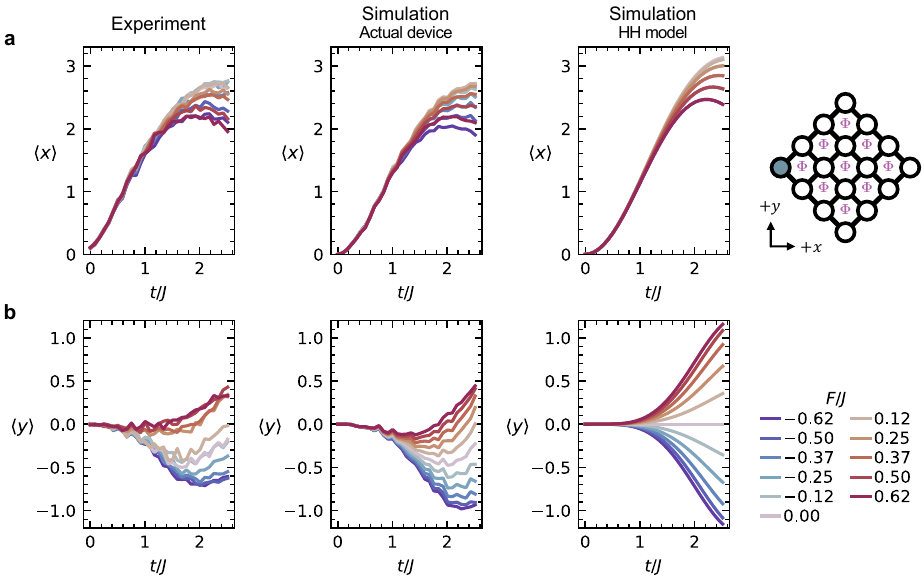}
\caption{\textbf{Position versus time} in the Hall effect experiment at synthetic magnetic flux $\pi/6$ per plaquette. \textbf{(a)} Longitudinal position. \textbf{(b)} Transverse position. Experimental data are shown at various synthetic electric field strengths and accompanied by simulations of the device and of the idealized Harper-Hofstadter model. Coordinate axes are drawn at right, and the initial position of the particle (teal site) is defined as point $(x,\ y) = (0,\ 0)$. Positions are written in units of unit cell length, e.g. the position of the rightmost site is $(x,\ y) = (3\sqrt{2},\ 0)$.} 
\label{sfig:hall_position}
\end{figure*}

\begin{figure*}[ht!]
\subfloat{\label{sfig:hall_coeff_E}}
\subfloat{\label{sfig:hall_coeff_B}}
\includegraphics{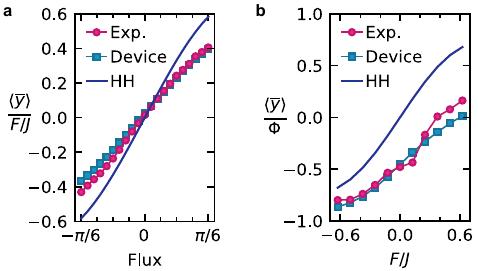}
\caption{\textbf{Hall coefficients}. \textbf{(a)} Average transverse deflection per electric field, shown as a function of magnetic flux, reproduced from Fig.~5c of the main text. \textbf{(b)} Average transverse deflection per magnetic flux, shown as a function of electric field.} 
\label{sfig:hall_coeff}
\end{figure*}

In Fig.~\ref{sfig:hall_position}, we present the average longitudinal and transverse position of a particle as a function of time as the particle travels through the $4\times 4$ array with synthetic magnetic flux per plaquette $\Phi_P=\pi/6$. Data is shown for various electric fields. An important observation is that the longitudinal position of the particle is roughly the same for electric fields $F$ and $-F$, yet the transverse deflection is not. This feature is not consistent with a description of the Hall effect in terms of a Lorentz force $\bm v \times \bm B$ for a classical velocity $\bm v$. In the main text, we presented coefficients representing the transverse deflection of a particle due to the Hall effect per electric field, which we compare to a Hall resistance (reproduced in Fig.~\ref{sfig:hall_coeff_E}). In Fig.~\ref{sfig:hall_coeff_B}, we present coefficients representing the same data per magnetic field. The offset between experimental results and results from an idealized Harper-Hofstadter model reflects the additional transverse deflection caused by inhomogeneity in the effective coupling strengths $J^{ij}$ across the lattice.

\begin{figure*}
\subfloat{\label{methods:hall_a}}
\subfloat{\label{methods:hall_b}}
\subfloat{\label{methods:hall_c}}
\subfloat{\label{methods:hall_d}}
\includegraphics{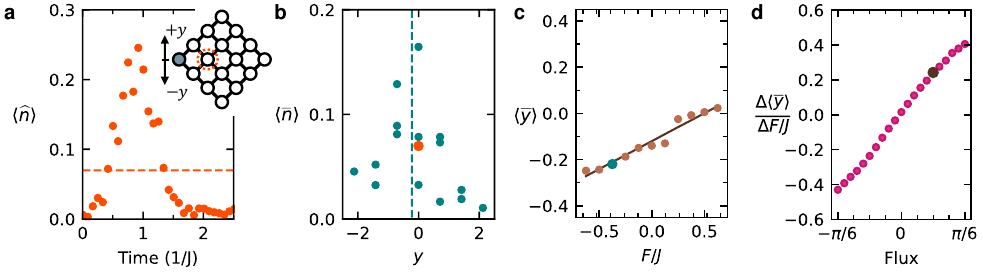}
\caption{\textbf{Analysis of Hall effect data}. \textbf{(a)} The population $\langle \hat n \rangle$ as a function of time. The data shown is for the site indicated by the orange circle in the inset measured at $F=-0.374J$ and $\Phi=\pi/12$. The time-averaged population $\langle \bar n \rangle$ of the site is indicated by the horizontal dashed line. \textbf{(b)} The value of $\langle \bar n \rangle$ for each site (at $F=-0.374J$, $\Phi=\pi/12$), shown as a function of each site's transverse position $y$. The orange point represents the site shown in the previous subpanel. The vertical teal line indicates the average $y$ position of the particle $\langle \bar y \rangle$. \textbf{(c)} The value of $\langle \bar y \rangle$ at $\Phi=\pi/12$ and different values of $F$. The teal point represents the value determined in the previous subpanel ($F=-0.374J$). The brown line displays a linear fit to these data; we define the Hall coefficient $\Delta \langle \bar{y}\rangle/\Delta F$ as its slope. \textbf{(d)} The Hall coefficient at different values of synthetic magnetic field. The brown point indicates the value determined in the previous subpanel ($\Phi=\pi/12$).} 
\label{methods:Hall_analysis}
\end{figure*}

\begin{figure*}
\includegraphics{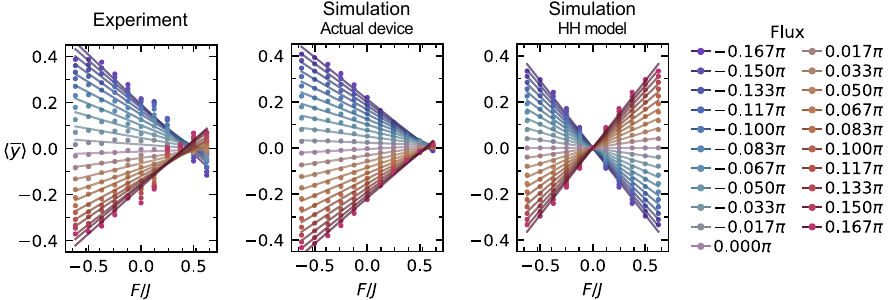}
\caption{\textbf{Extracting Hall coefficients}. The average transverse deflection data shown in Fig.~5b of the main text are reproduced as a function of the synthetic electric field (circles). Linear fits (lines) are shown for each synthetic magnetic flux value, and are used to determine the Hall coefficients shown in Fig.~5c of the main text.} 
\label{ext:hall_fitting}
\end{figure*}

In Fig.~\ref{methods:Hall_analysis}, we describe our analysis of Hall effect data and the extraction of Hall coefficients. At each value of the synthetic electric and magnetic fields, we measure the population $\langle \hat n_i \rangle$ of each site $i$ as a function of time. The time average of these data yield the average populations $\langle \bar n_i \rangle$ (Fig.~\ref{methods:hall_a}). The average position of the particle is given by the weighted average $\langle \bar y \rangle = \sum_i \langle \bar n_i \rangle y_i$ where $y_i$ is the transverse coordinate of site $i$ (Fig.~\ref{methods:hall_b}). The value of $\langle \bar y \rangle$ is determined at each value of the synthetic electric field; the Hall coefficient $\Delta\langle \bar y \rangle/\Delta F$ is determined as the slope of a linear fit to these data (Fig.~\ref{methods:hall_c}). This process is repeated to determine $\Delta\langle \bar y \rangle/\Delta F$ at each value of the synthetic magnetic field (Fig.~\ref{methods:hall_d}). All linear fits used to determine the Hall coefficients shown in Fig.~5c of the main text are shown in Fig.~\ref{ext:hall_fitting}.

\clearpage
\section{Simulation details and extended simulation results}

\subsection{Description of simulations}

In the main text, we present simulations of two models: a model of the actual device as it is operated in the experiment, and the ideal Harper-Hofstadter model. The two models differ in two primary ways. First, in the former, on-site energies are modulated and coupling is parametrically induced, whereas in the latter, all sites are resonant. Simulations of the ideal Harper-Hofstadter model, therefore, lack rotating contributions to the dynamics (see Section~\ref{sec:coupling_derivation}). Second, in the former, the bare couplings $J_0^{ij}$ between nearest and next-nearest neighbors, as experimentally determined, are used, whereas the latter assumes idealized couplings (uniform nearest-neighbor coupling and no next-nearest-neighbor coupling).

To further elucidate these differences, we present four models in this section. 
In all models, the average effective coupling rate is $J/2\pi=\SI{2.0}{MHz}$, and $J_0/2\pi=\SI{5.8}{MHz}$ is the average bare coupling of the device.

\noindent 1. We consider a model including parametric coupling and the real values of $J_0^{ij}$. The Hamiltonian of this model is
\begin{equation}\label{eq:sup_H1}
    \hat{H}_1/\hbar = \sum_{i} \big(\omega_i + \Omega_i\sin(\delta_{i} t+\phi_{i})\big)n_i +\sum_{\langle i, j \rangle} J_0^{ij} \left(\hat{a}_i^\dag \hat{a}_j+\hat{a}_j^\dag \hat{a}_i \right),
\end{equation}
and this model is our most complete model of the experimental operation of the actual device.

\noindent 2. We consider a model including the parametric coupling scheme used to operate the actual device, but with idealized couplings.
\begin{equation}\label{eq:sup_H2}
    \hat{H}_2/\hbar = \sum_{i} \big(\omega_i + \Omega_i\sin(\delta_{i} t+\phi_{i})\big)n_i +J_0\sum_{\langle i, j \rangle} \left(\hat{a}_i^\dag \hat{a}_j+\hat{a}_j^\dag \hat{a}_i \right),
\end{equation}

\noindent 3. We consider the Harper-Hofstadter model with the real values of $J_0^{ij}$.
The Hamiltonian of this model is
\begin{equation}\label{eq:sup_H3}
    \hat{H}_3/\hbar = \sum_{\langle i, j \rangle} J_{ij} \left(e^{i\phi_{ij}}\hat{a}_i^\dag \hat{a}_j + e^{-i\phi_{ij}}\hat{a}_i \hat{a}_j^\dag \right),
\end{equation}
where $J_{ij} = \frac{J}{J_0}J_0^{ij}$.

\noindent 4. We consider the Harper-Hofstadter model with idealized couplings. The Hamiltonian of this model is
\begin{equation}\label{eq:sup_H4}
    \hat{H}_4/\hbar = J \sum_{\langle i, j \rangle} \left(e^{i\phi_{ij}}\hat{a}_i^\dag \hat{a}_j + e^{-i\phi_{ij}}\hat{a}_i \hat{a}_j^\dag \right).
\end{equation}
Simulations using models $\hat{H}_1$ and $\hat{H}_4$ are presented in the main text (``simulation: actual device" and ``simulation: HH model", respectively). 

For the models $\hat{H}_1$ and $\hat{H}_2$ with parametric coupling, the modulation amplitudes $\Omega_i$ were chosen by replicating the tuneup procedure discussed in Section~\ref{sec:tuneup_amp} in simulations. As each site is modeled as a two-level system (rather than as a transmon qubit), compensation for DC frequency shifts is unneeded.

For models $\hat{H}_1$ and $\hat{H}_2$, when simulating experiments with a synthetic electric field, the electric field was generated in the same manner as in the experiment: a time varying phase $\varphi t$ was added to each modulation tone, which is equivalent to shifting the modulation frequencies $\delta_i \rightarrow \delta_i+\varphi$. For models $\hat{H}_3$ and $\hat{H}_4$, we include an idealized representation of the electric field by adding the linear potential term
\begin{equation}\label{eq:F}
    \hat{H}_F/\hbar = \sum_{i} F x_i \hat{n}_i,
\end{equation}
where $F$ is the field strength per site and $x_i$ is the position of site $i$ along the $x$-axis in units of the lattice constant.

\subsection{Comparison between different models}

\begin{figure*}[ht!]
\includegraphics{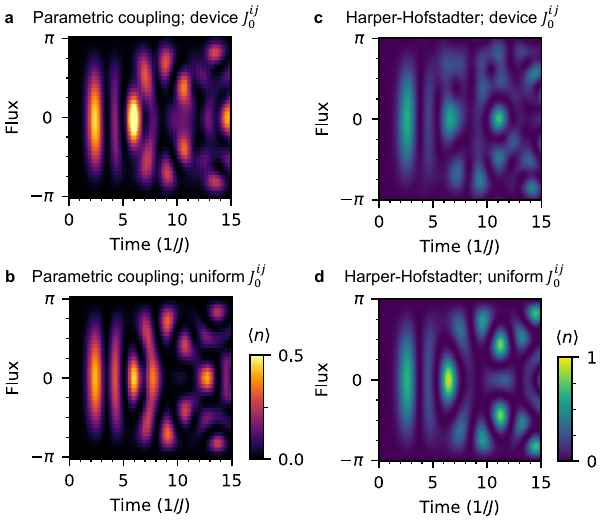}
\caption{\textbf{Aharonov-Bohm interference: comparing simulations using different models} of the experiment in an $8$-site ring presented in Fig.~2b of the main text. \textbf{(a-d)} Simulations using the models described in Equations~(\ref{eq:sup_H1}-\ref{eq:sup_H4}), respectively.}
\label{sfig:3x3_sim_comparison}
\end{figure*}

In Fig.~\ref{sfig:3x3_sim_comparison}, we present simulations of the experiment presented in Fig.~2b of the main text---Aharonov-Bohm interference in an $8$-site ring---using models $\hat{H}_1$ through $\hat{H}_4$. The four simulations share their primary features. A notable difference is that the interference pattern under $\hat{H}_3$ is less regular than the pattern under $\hat{H}_1$, likely reflecting that the parametric coupling scheme partially corrects for disorder in the bare coupling rates, as discussed in Section~\ref{sec:effective_j}.

\begin{figure*}[ht!]
\includegraphics{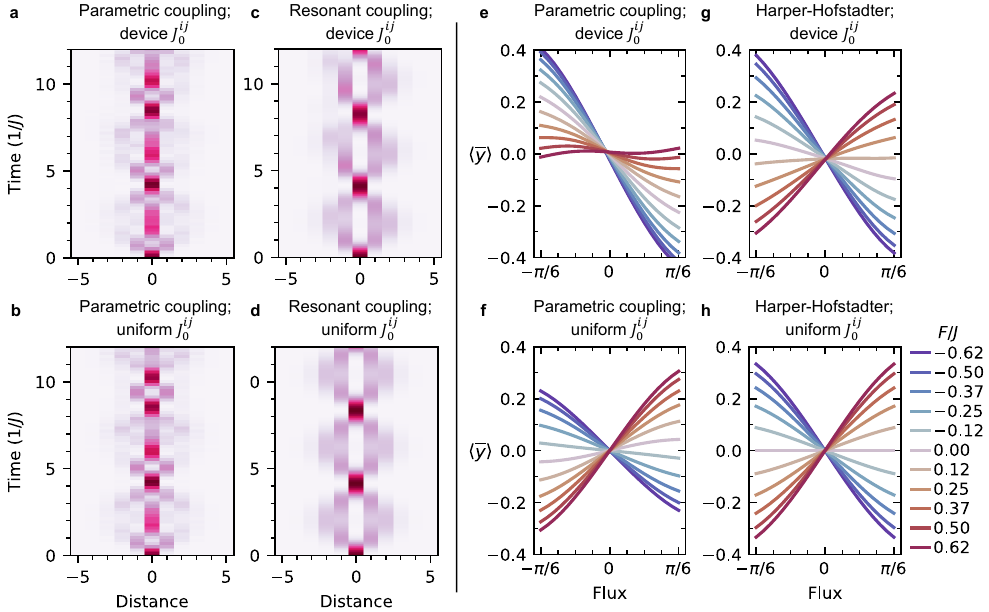}
\caption{\textbf{Localization in a synthetic electric field and the Hall effect: comparing simulations using different models}. \textbf{(a-d)} Simulations using the models described in Equations~(\ref{eq:sup_H1}-\ref{eq:sup_H4}), respectively, for the experiment presented in Fig.~4 of the main text. \textbf{(e-h)} Simulations using the models described in Equations~(\ref{eq:sup_H1}-\ref{eq:sup_H4}), respectively, for the experiment presented in Fig.~5 of the main text.}
\label{sfig:hall_sim_comparison}
\end{figure*}

In Fig.~\ref{sfig:hall_sim_comparison}(a-d), we present simulations of the experiment presented in Fig.~4 of the main text---Wannier-Stark localization in a 1D chain from a synthetic electric field---using models $\hat{H}_1$ through $\hat{H}_4$. We find that differences between the idealized model and the device behavior predominantly come from the nonidealities in the parametric coupling, that is, the nonstationary terms in the rotating frame Hamiltonian Eq.~(\ref{eq:full_HR}). Interestingly, coupling rate disorder has little impact on the dynamics of this experiment, which we interpret to be a consequence of using a 1D chain: in a 2D system, coupling rate disorder changes the interference between different trajectories of a particle around plaquettes. In a 1D system, there are not multiple non-overlapping trajectories between sites, so coupling rate disorder affects the dynamics comparatively less.

In Fig.~\ref{sfig:hall_sim_comparison}(e-h), we present simulations of the experiment presented in Fig.~5 of the main text---Hall effect in the $4\times 4$ lattice---using models $\hat{H}_1$ through $\hat{H}_4$. While all four simulations feature increasing Hall coefficient (slope of $\langle \bar{y} \rangle$ as a function of flux) with increasing synthetic electric field, they have different offsets (different hall coefficients at $F=0$). Simulations under $\hat{H}_3$ yield results most similar to those under $\hat{H}_1$ (which, as shown in Fig.~5, match experiment), suggesting that coupling rate disorder is the leading cause of the offset.

\subsection{Simulations with decoherence}

\begin{figure*}[ht!]
\includegraphics{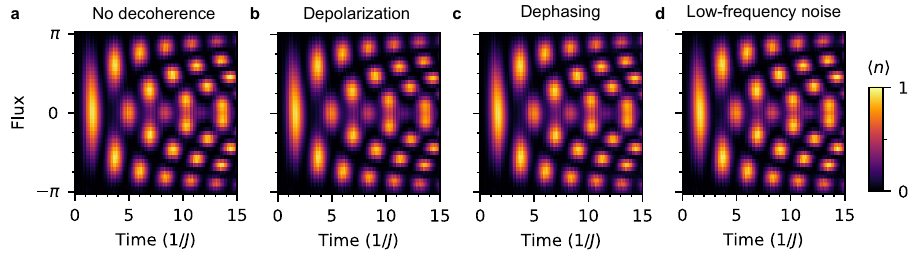}
\caption{\textbf{Simulations of Aharonov-Bohm interference with decoherence}. \textbf{(a)} The simulations of the actual device operation presented in Fig.~2a of the main text ($2\times 2$ plaquette) is reproduced. \textbf{(b)} Simulation including depolarization. \textbf{(c)} Simulation including pure dephasing, with rates determined by Hahn echo. \textbf{(d)} Simulation including quasi-static on-site disorder, with rates determined by Ramsey interferometry, to simulate low-frequency flux noise.}
\label{sfig:2x2_decoherence}
\end{figure*}

\begin{figure*}[ht!]
\includegraphics{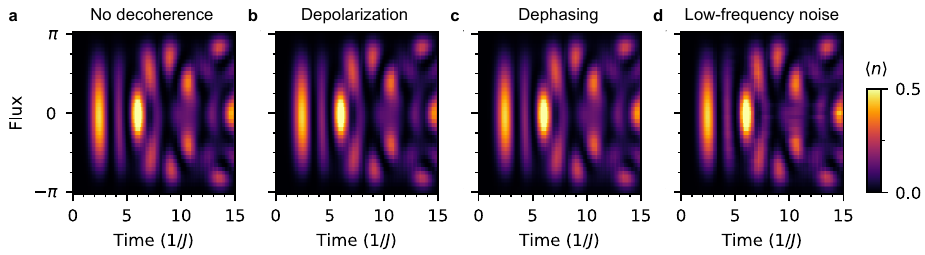}
\caption{\textbf{Simulations of Aharonov-Bohm interference with decoherence}. \textbf{(a)} The simulation of the actual device operation presented in Fig.~2b of the main text ($8$-site ring) is reproduced. \textbf{(b)} Simulation including depolarization. \textbf{(c)} Simulation including pure dephasing, with rates determined by Hahn echo. \textbf{(d)} Simulation including quasi-static on-site disorder, with rates determined by Ramsey interferometry, to simulate low-frequency flux noise.}
\label{sfig:3x3_decoherence}
\end{figure*}

In Fig.~\ref{sfig:2x2_decoherence}, we present simulations with decoherence of the $2\times 2$ Aharonov-Bohm interferometry experiment discussed in Fig.~2a of the main text. For these simulations, we use the full model of the system Equation~(\ref{eq:sup_H1}). In the first panel, we reproduce the decoherence-free simulation. In the second panel, we include depolarization using the measured $T_1$ of each qubit. In the third panel, we include pure dephasing with rates determined by Hahn echo measurements; these simulations describe the effect of high-frequency noise. In the fourth panel, we present simulations averaged over many instantiations of random on-site disorder, with the disorder amplitude determined by the decoherence time in Ramsey interferometry. These simulations describe the effect of low-frequency flux noise, which shifts the qubits' frequencies but can be assumed to be static during each repetition of the measurement sequence. Because the measurement time ($\SI{1.2}{\mu s}$) is short compared to the depolarization and dephasing rates, the decoherence channels do not have a dramatic effect on the measurement. In Fig.~\ref{sfig:3x3_decoherence}, we present simulations with decoherence of the $8$-site Aharonov-Bohm interferometry experiment discussed in Fig.~2b of the main text.

\clearpage
\section{Details about Hall velocity in idealized model}
We consider a Harper-Hofstadter type tight-binding model on a square lattice as our underlying system. For an electric field $  E_x \hat{x}+E_y \hat{y} $, magnetic field $ B_z \hat{z}$ corresponding to dimensionless flux $\Phi = qa^2 B_z/\hbar $, unit cell length $a$, and charge $q$, the Hamiltonian in the Landau gauge is given by
\begin{equation}\label{eq:HH}
	\widehat{H} /\hbar
	=
	-J \sum_{\mathbf{r}}
	\left (
	\hat{a}_{\mathbf{r}}^\dagger \hat{a}_{\mathbf{r}+\hat{y}} 
	+ \ee^{-\ii \Phi y} \hat{a}_{\mathbf{r}}^\dagger \hat{a}_{\mathbf{r}+\hat{x}}  
	+H.c.
	\right )
	+
	q\sum_{\mathbf{r}} \left (xE_x + yE_y\right )\hat{n}_{\mathbf{r}},
\end{equation}
noting that the synthetic electric field $F$ described in the experiment is equivalent to $q|\textbf{E}|/a$. We assume that $ E_x $ is much smaller than the hopping strength $ J $ and that $ B_z $ is also appropriately weak. Then we can apply semi-classical transport results to our problem. 
For position ${\bm r}$ and quasimomentum ${\bm k}$, the wave packet dynamics are given by the following semi-classical equations of motion: \cite{XCN101103}
\begin{align}\label{eq:eom1}
	\dot{\mathbf{r}} &= 
	\frac{1}{\hbar} \frac{\pt \veps_M(\mathbf{k})}{\pt \mathbf{k}} 
	- \dot{\mathbf{k}}\times \mathbf{\Om}(\mathbf{k})
	\\
	\hbar \dot{\mathbf{k}} &=
	q\mathbf{E} + q\dot{\mathbf{r}}\times \mathbf{B}
\end{align}
where $  \veps_M(\mathbf{k}) =  \veps_0(\mathbf{k}) -\mathbf{B}\cdot \mathbf{m}(\mathbf{k})$ with $ \mathbf{m}(\mathbf{k}) $ being the orbital angular moment of the wave packet.

\subsection{External \texorpdfstring{$ E $}{E} and \texorpdfstring{$ B $}{B} fields via semiclassical dynamics}

We can treat the phases in the hopping terms as coming from a magnetic field applied on top of a bare hopping band. The equations of motion above reduce to
\begin{align}
	\hbar \dot{\mathbf{r}} &= 
	\frac{\pt \veps(\mathbf{k})}{\pt \mathbf{k}}\label{eq:rdot1}
	\\
	\frac{\hbar}{q} \dot{\mathbf{k}} &=
	\mathbf{E} + \dot{\mathbf{r}}\times \mathbf{B}\label{eq:kdot1}
\end{align}
where we dropped the subscript $ M $ in the first equation since for a trivial band, the angular moment of the wave packet should vanish. 
Substituting \eqn{eq:rdot1} in \eqref{eq:kdot1}, we get
\begin{equation}\label{eq:rdot1sub}
	\frac{\hbar}{q} \dot{\mathbf{k}} =
	\mathbf{E} + \frac{1}{\hbar}  \frac{\pt \veps(\mathbf{k})}{\pt \mathbf{k}} \times \mathbf{B}\
\end{equation}

In the experiments, we have open boundary conditions (OBCs), so the ordinary plane wave solutions are not good eigenstates anymore. Let us discuss how to correctly parametrize the eigenstates of the hopping Hamiltonian with OBCs.
Consider a square lattice that has $ N_x \times N_y$ sites. 
For computational convienience, in this section we use coordinates where $x, y\in [0, aN_{x,y}]$, i.e. coordinates rotated by $\pi/4$ with respect to the coordinates chosen in the main text. 
Then the particle is initialized at the corner $x=y=a$, the longitudinal electric field is in the direction $\hat x+\hat y$, and the transverse direction is $\hat x - \hat y$.

\begin{figure*}[t]
\subfloat{\label{sfig:dispersion-with-obc}}
\subfloat{\label{sfig:initialcondition}}
\subfloat{\label{sfig:longitudinal_velocity}}
\subfloat{\label{sfig:Hall_velocity}}
\includegraphics{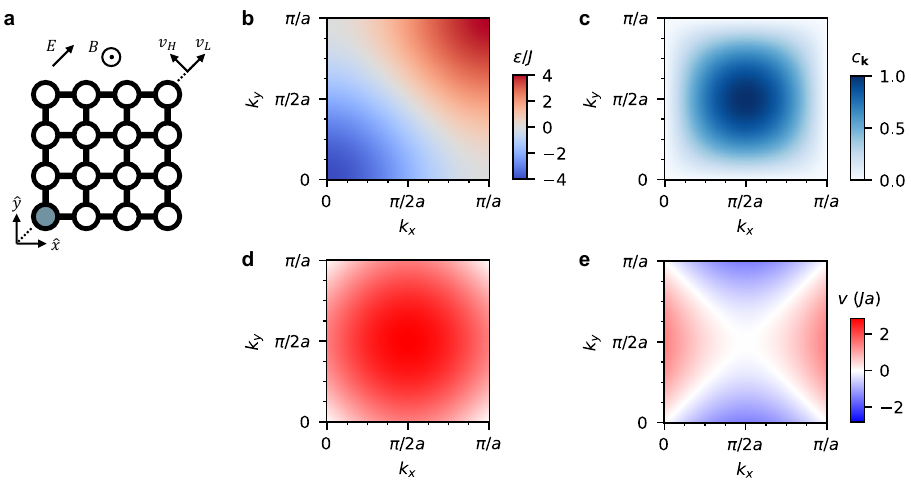}
\caption{\textbf{Tight-binding model in momentum space}. \textbf{(a)} A diagram of the experiment in real space, with the coordinates introduced in this section. A $4\times 4$ lattice is shown. The particle is initialized in a position state at lower left (teal shaded site). The longitudinal and Hall velocities $\bm v_{L,\, H}$ are parallel and transverse to the electric field, respectively, and the magnetic field is out of plane. \textbf{(b)} Energy eigenvalues $ \veps(k_x,k_y) $ labeling the eigenstates as in \eqn{eq:ansatz}. \textbf{(c)} A plot of $ \sin k_x a \sin k_y a $ representing the overlap of the initial wave function with the eigenstates labeled by~$ \mathbf{k} $. \textbf{(d)} The longitudinal velocity (velocity in the direction $\hat x + \hat y$). \textbf{(e)} The Hall velocity (velocity in the direction $\hat x - \hat y$). Note the two-fold mirror antisymmetry about ${\bm k} = (\pi/2a, \pi/2a)$ along the lines $k_x=k_y$ and $k_x=-k_y$. } 
\label{sfig:hall_kspace}
\end{figure*}

As the wave function $ \psi(x,y) $ must vanish outside the lattice -- which consists of the sites with $ x/a=1,2,\dots N_x $ and $ y/a=1,2,\dots N_y $ -- we have the constraints
\begin{equation}\label{eq:obc}
	\psi(aN_x+a,y) = 0, \ \psi(0,y) = 0,\ \psi(x,aN_y+a) = 0, \ \psi(x,0) = 0.
\end{equation}
This motivates the following ansatz 
\begin{equation}\label{eq:ansatz}
	\psi_{k_x,k_y}(x,y)  = A_{k_x,k_y}\sin (k_x x)\sin(k_y y) \,,
\end{equation}
with $ k_x a(N_x+1) \in \pi \mathbb Z $ and $ k_y a(N_y+1) \in \pi \mathbb Z $.
We emphasize that, due to the OBCs, the momenta are defined modulo $\pi/a$, whereas in problems with periodic boundary conditions, momenta are defined modulo $2\pi/a$ (the Brillouin zones of lattices are typically presented with momenta spanning from $-\pi/a$ to $\pi/a$).
The bare hopping Hamiltonian with open boundary conditions is
\begin{equation*}
	H /\hbar= -J \sum_{x,y\in \text{int}}\left [ a_{x+1,y}^\dagger a_{x,y}+ a_{x,y+1}^\dagger a_{x,y} +h.c.\right ] + H_{\text{bdy}},
\end{equation*}
where ``int" refers to the sites that are in the interior of the lattice and $ H_{\text{bdy}} $ is an appropriately modified set of terms on the boundary that respect the open boundary conditions, \eg near the left boundary at $ x=0 $, there are no leftward hop terms. By explicitly applying the Hamiltonian on the ansatz wavefunction we find
\begin{align*}\label{eq:Hpsi}
	H  \psi_{k_x,k_y}(x,y) /\hbar
	&=
	-J\left [ \psi_{k_x,k_y}(x+1,y) +\psi_{k_x,k_y}(x-1,y) + \psi_{k_x,k_y}(x,y+1) + \psi_{k_x,k_y}(x,y-1) \right ]\\
	&= -2J\left (\cos k_xa +\cos k_ya \right )  \psi_{k_x,k_y}(x,y) \equiv  \veps(k_x,k_y) \psi_{k_x,k_y}(x,y) \,,
\end{align*}
which confirms for us the validity of our ansatz. The eigenenergies are given by $ \veps(k_x,k_y) $. 
% \begin{figure}[t]
% 	\centering
% 	\includegraphics[width=0.4\linewidth]{"sfig_dispersion_obc"}
% 	\caption{Energy eigenvalues $ \veps(k_x,k_y) $ labeling the eigenstates as in \eqn{eq:ansatz}.}
% 	\label{fig:dispersion-with-obc}
% \end{figure}
We note that the constraints \eqref{eq:obc} restricts the allowed values of $ k_x $ and $ k_y $ to the following linearly independent set of modes:
\begin{equation}\label{eq:allowed k}
	k_x \in \left \{\frac{\ell_x\pi/a}{N_x+1} \, \Big| \, \ell_x=1,2,\dots N_x\right \}, 
	\quad
	k_y \in \left \{\frac{\ell_y\pi/a}{N_y+1} \, \Big| \, \ell_y=1,2,\dots N_y\right \}
\end{equation}
The energy eigenvalues are plotted in Fig. \ref{sfig:dispersion-with-obc} Interestingly we note the symmetry of the ``band" upon reflection about the line $ k_x=k_y $. Moreover, note that for every eigenstate with energy $ E $, there is another eigenstate with energy $ -E $. The corresponding $ \mathbf{k} $ values are related by reflection about the line $ k_x+k_y = \pi/a $. 

The initial condition in the experiment is a singly occupied site in the corner of the lattice, say at $ (x,y) = (a,a) $. The overlap of this state with the eigenstates can be obtained by an inverse discrete sine transform over the modes mentioned in \eqn{eq:allowed k}. So, from the wavefunction we started with,
\begin{equation*}
\psi_0(x,y) = \delta_{x,a}\delta_{y,a} 
\equiv 
\sum_{\ell_x=1}^{N_x} 
\sum_{\ell_y=1}^{N_y} c_{\ell_x,\, \ell_y}
\sin \frac{\pi \ell_x x /a}{N_x+1}
\sin \frac{\pi \ell_y y /a}{N_y+1} \,,
\end{equation*}
we can get the Fourier sine mode coefficients
\begin{equation}\label{eq:iDST}
	\begin{split}
		c_{\ell_x,\, \ell_y} 
		&= 
		\frac{2}{N_x+1}\frac{2}{N_y+1}
		\sum_{x=1}^{N_x}
		\sum_{y=1}^{N_y}
		\psi_0(x,y) \sin \frac{\pi \ell_x x/a }{N_x+1}
		\sin \frac{\pi \ell_y y/a }{N_y+1}\\
		&=
		\frac{4}{(N_x+1)(N_y+1)} \sin \frac{\pi \ell_x  }{N_x+1}
		\sin \frac{\pi \ell_y  }{N_y+1} = \frac{4}{(N_x+1)(N_y+1)} \sin k_x a
		\sin k_y a \,,
	\end{split}
\end{equation}
which is distributed symmetrically about $ (k_x a,k_y a)\approx \left (\frac{\pi}{2},\frac{\pi}{2}\right ) $. In particular, the initial momentum distribution has mirror symmetries about the lines $x=y$ and $x=-y$. This is shown in Fig. \ref{sfig:initialcondition}.
% \begin{figure}[t]
% 	\centering
% 	\includegraphics[width=0.4\linewidth]{"sfig_initialcondition_overlap"}
% 	\caption{A plot of $ \sin k_x a \sin k_y a $ representing the overlap of the initial wave function with the eigenstates labeled by $ \mathbf{k} $.}
% 	\label{fig:initialcondition-overlap}
% \end{figure}
%We can check this by explicit computation,
Therefore, we find that the eigenstates with the largest overlap with our localized initial state are symmetrically around $ (k_xa,k_ya) \approx \left ( \frac{\pi}{2},  \frac{\pi}{2} \right )$. 
Given this initial condition, let us now solve the semiclassical equations of motion \eqref{eq:rdot1} and \eqref{eq:kdot1}. From the band dispersion,
\begin{equation*}
	\veps(\mathbf{k}) = -2\hbar J \left (\cos k_x a + \cos k_y a\right )
\end{equation*}
we get the velocity
\begin{equation}\label{eq:vel}
	 \frac{\pt \veps(\mathbf{k})}{\pt \mathbf{k}} = 
	2\hbar J a \left (\sin (k_x a) \ \hat{x} +  \sin (k_y a) \ \hat{y}\right )\,.
\end{equation}
This can be substituted into \eqn{eq:rdot1sub} to get
\begin{align*}
	\frac{\hbar}{q} \dot{k_x}
	&= E_x + 2 J B_z \sin k_y a \\
	\frac{\hbar}{q} \dot{k_y}
	&= E_y - 2 J B_z \sin k_x a
\end{align*}

\begin{figure*}[t]
\includegraphics{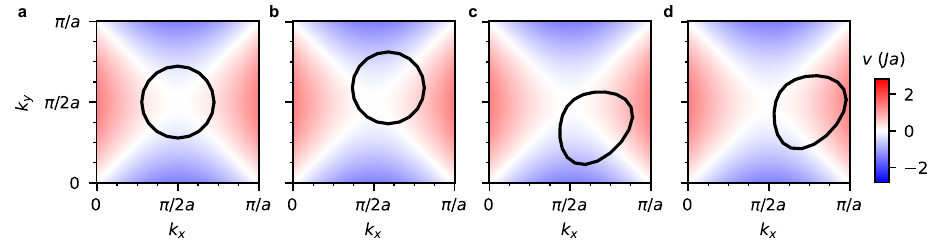}
\caption{\textbf{Visualizing the evolution of the wavepacket in momentum space}. The transverse velocity $v_H$ throughout the Brillouin zone is displayed in the backgrounds. \textbf{(a)} An initial contour is drawn symmetrically about ${\bm k} = (\pi/2a, \pi/2a)$, representing the initial wavepacket. \textbf{(b)} The contour after evolving in an electric field $E=0.4J/a$ and zero magnetic field for time $t=1/J$. Due to the mirror antisymmetry of $v_H$ along the line $k_x=k_y$, there is no transverse deflection. \textbf{(c)} The contour after evolving in zero electric field and $\pi/12$ magnetic flux per plaquette. Due to the mirror antisymmetry of $v_H$ along the line $k_x=-k_y$, there is no transverse deflection. \textbf{(d)} The contour after evolving in an electric field $E=0.4J/a$ and $\pi/12$ magnetic flux per plaquette. When both $E$ and $B$ are nonzero, the wavepacket evolution leaves the axes of mirror antisymmetry, permitting transverse deflection.} 
\label{sfig:hall_kspace}
\end{figure*}

For simplicity, let us work in units where $\hbar=q=a=1$ and linearize near $  (k_x,k_y) \approx \left ( \frac{\pi}{2},  \frac{\pi}{2} \right ) $. Defining $ \kappa_x = k_x-\frac{\pi}{2} $ and $ \kappa_y = k_y-\frac{\pi}{2}$, we have
\begin{align*}
	\dot{\kappa}_x
	&= E_x + B_z 2J \sin \left (\frac{\pi}{2}+\kappa_y\right ) = E_x + B_z 2J \cos \left (\kappa_y\right ) \approx E_x+2JB_z +\cO(\mathbf{\kappa}^2)\\
	\dot{\kappa}_y
	&=E_y - B_z 2J \sin  \left (\frac{\pi}{2}+\kappa_x\right ) 
	=E_y -B_z 2J \cos  \left (\kappa_x\right ) 
	\approx E_y-2JB_z +\cO(\mathbf{\kappa}^2)
\end{align*}
So at linear order in $ \bm{\kappa} $, the solution to these equations is given as
\begin{equation}\label{eq:solq}
	\kappa_x = \kappa_{x,0} + (E_x+2JB_z) t \,,
	\quad 
	\kappa_y = \kappa_{y,0} + (E_y-2JB_z)  t \,,
\end{equation}
where $\bm{\kappa}_0$ is the momentum at time $t=0$, which then implies, via  \eqn{eq:vel},
\begin{equation}\label{eq:rsol}
	\begin{split}
		\dot{x} 
%		= 2J \sin k_x(t) 
		&= 2J \cos\left (\kappa_{x,0} +(E_x+2JB_z) t \right ) \approx  
		2J \big[ \cos((E_x+2JB_z)t) + \kappa_{x,0} \sin((E_y+2JB_z)t)   \big ]\\
		\dot{y} 
%		= 2J \sin k_y(t) 
		&
		=2J\cos  \left (  \kappa_{y,0} +(E_y-2JB_z) t \right )
		\approx
		2J \big [ \cos((E_y-2JB_z)t) + \kappa_{y,0} \sin((E_y-2JB_z)t)   \big ]\
	\end{split}
\end{equation}
In particular, let us set $ E_x=E_y=E_0/\sqrt{2} $ and compute the velocity along the line $ x=y $:
\begin{equation}\label{eq:diagvel}
	\frac{\dot{x}+\dot{y}}{\sqrt 2} =\sqrt 2 J\left[ 2\cos\left(\frac{E_0}{\sqrt{2}}t \right)\cos(2JB_zt) +  \kappa_{x,0} \sin\left(\left(\frac{E_0}{\sqrt{2}}+2JB_z\right)t \right)  + \kappa_{y,0} \sin\left(\left(\frac{E_0}{\sqrt{2}}-2JB_z\right)t\right) \right]
\end{equation}
Similarly, the velocity perpendicular to the line $ x=y $ is given by:
\begin{equation}\label{eq:crossdiagvel}
	\frac{\dot{x}-\dot{y}}{\sqrt 2} = \sqrt 2J\left[ 2\sin\left(\frac{E_0}{\sqrt{2}}t\right)\sin(2JB_zt) +  \kappa_{x,0} \sin\left(\left(\frac{E_0}{\sqrt{2}}+2JB_z\right)t \right)  + \kappa_{y,0} \sin\left(\left(\frac{E_0}{\sqrt{2}}-2JB_z\right)t\right) \right]
\end{equation}
From this solution, we can see that when $ B_z=0 $, the Hall velocity must vanish when we take the average over a set of initial momenta $\bm{\kappa}_0$ that are symmetrically distributed about $ \bm{\kappa}=0 $ (which is the initial condition as discussed above), \ie
\begin{equation}\label{eq:perp vel no B field}
	v_H\rvert_{B_z=0} = \left \<\frac{\dot{x}-\dot{y}}{\sqrt{2} } \right \>\Big \rvert_{B_z=0} =\sqrt{2}J \left ( \<\kappa_{x,0}\>  + \<\kappa_{y,0}\>\right ) \sin(E_0t/\sqrt{2}) = 0,
\end{equation}
where the brackets indicate an average across the momenta comprising the initial wavepacket with coefficients given by Eq.~(\ref{eq:iDST}). Similarly, the Hall velocity also vanishes when $E_0=0$:
\begin{equation}\label{eq:perp vel no E field}
	v_H\rvert_{E_0=0} = \left \<\frac{\dot{x}-\dot{y}}{\sqrt{2} } \right \>\Big \rvert_{E_0=0} =\sqrt{2}J \left ( \<\kappa_{x,0}\>  - \<\kappa_{y,0}\>\right ) \sin(2JB_zt) = 0.
\end{equation}

On the other hand, when $ B_z\neq 0 $ and $ E_0\neq 0 $, we have a non-zero Hall response
\begin{equation}\label{eq:perp vel with B field}
	v_H = \left \<\frac{\dot{x}-\dot{y}}{\sqrt{2} } \right \> = 2\sqrt{2}J\sin(E_0t/\sqrt{2})\sin(2JB_zt)
\end{equation}
which is, appropriately, odd under $ B_z\to -B_z $.

\begin{figure*}[ht!]
\includegraphics{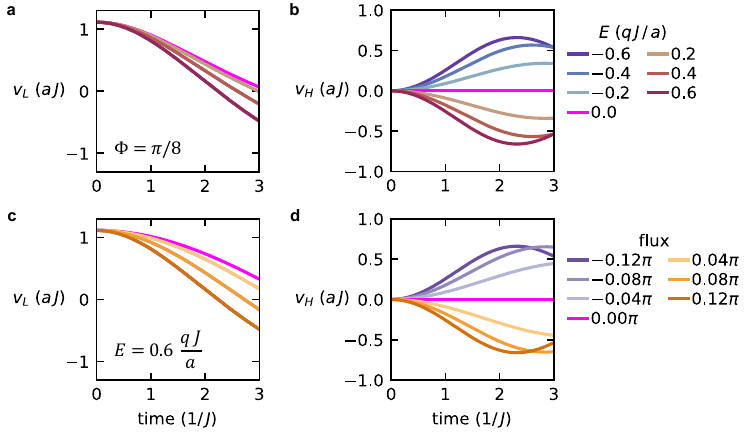}
\caption{\textbf{Velocity dynamics obtained from a semiclassical model} after initializing a particle in a corner of a $20\times 20$ site lattice. $\textbf{(a)}$ Longitudinal velocity at dimensionless flux $\pi/8$ per plaquette and various electric fields. $\textbf{(b)}$ Corresponding Hall velocity. Results at zero electric field are highlighted in magenta. $\textbf{(c)}$ Longitudinal velocity at electric field $0.6qJ/a$ and various magnetic fields. $\textbf{(d)}$ Corresponding Hall velocity. Results at zero magnetic field are highlighted in magenta.} 
\label{sfig:hall_trajectory}
\end{figure*}

If we were not to linearize in $ \bm{\kappa} $, we would have the fully non-linear coupled differential equations:
\begin{align*}
	\dot{\kappa}_x
	&= E_x + B_z 2J \cos \kappa_y
	\\
	\dot{\kappa}_y
	& 
	=E_y - B_z 2J \cos  \kappa_x
\end{align*}
along with the relation to real space velocity, \eqref{eq:vel},
\begin{equation*}
	(\dot{x}, \dot{y}) = 2J \left (\cos \kappa_x(t), \cos \kappa_y(t)\right )
\end{equation*}
from which we can find the velocity perpendicular to the $ x=y $ line given by 
\begin{equation}\label{eq:vH general}
	v_H = \left \< \frac{\dot{x}-\dot{y}}{\sqrt{2}} \right \>
	=2J\left \<\cos \kappa_x(t) - \cos \kappa_y(t) \right \>\,.
\end{equation}
By numerically integrating the coupled nonlinear ODEs, we can show that this quantity vanishes if either the electric field along the diagonal, $ E_0 $, or the magnetic field $ B_z $ vanishes. The results are shown in Fig. \ref{sfig:hall_trajectory}.

\subsection{The role of disorder}

According to Eqs.~(\ref{eq:perp vel no B field}, \ref{eq:perp vel no E field}), in the case of zero disorder, there is zero transverse deflection when the magnetic or electric field is zero due to the symmetry of the initial wavepacket.
In the presence of disorder in the coupling strengths, momentum is not a good quantum number and the symmetry of the initial wavepacket about $\bm{\kappa}=0$ breaks down.
As seen in the experimental results, when the electric field is zero, the particle deflects transversely as long as the magnetic field is nonzero.

Interestingly, we observe that when the magnetic field is zero (but the electric field is finite), the particle does not deflect transversely even with disorder.
To understand this, we note that the Hall effect requires broken inversion symmetry and broken time-reversal symmetry. 
Coupling strength disorder breaks inversion symmetry but preserves time-reversal symmetry. 
To observe a Hall effect, therefore, either an electric field or disorder is sufficient to break inversion symmetry, but coupling strength disorder cannot replace role of the magnetic field in breaking time-reversal symmetry.

\bibliography{refs}